\documentclass[fleqn,usenatbib]{mnras}

\usepackage{newtxtext,newtxmath}
\usepackage{chemformula}
\usepackage{booktabs}
\usepackage{pdflscape}

\usepackage[T1]{fontenc}

\usepackage{graphicx}	
\usepackage{amsmath}	
\usepackage{siunitx}
\usepackage[version=4]{mhchem}
\usepackage{xspace}
\usepackage{orcidlink}
\usepackage{natbib}

\newcommand{\degree}{$^{\circ}$}

\newcommand{\mystar}{KELT-7\xspace}
\newcommand{\myplanet}{KELT-7\,b\xspace}
\newcommand{\eureka}{\texttt{Eureka!}\xspace}
\newcommand{\tiberius}{\texttt{Tiberius}\xspace}
\newcommand{\exoticjedi}{\texttt{ExoTiC-JEDI}\xspace}
\newcommand{\pRT}{\texttt{pRT}\xspace}
\newcommand{\poseidon}{\texttt{POSEIDON}\xspace}
\newcommand{\nemesispy}{\texttt{NEMESISPY}\xspace}
\newcommand{\nemesis}{\texttt{NEMESIS}\xspace}

\defcitealias{MayMacDonald2023ApJ}{May \& MacDonald et al. 2023}

\title[JWST NIRSpec transmission spectrum of \myplanet]{BOWIE-ALIGN: Weak spectral features in KELT-7b's JWST NIRSpec/G395H transmission spectrum imply a high cloud deck or a low-metallicity atmosphere}

\author[E. Ahrer et al.]{Eva-Maria Ahrer$^{\orcidlink{0000-0003-0973-8426},1}$\thanks{E-mail: ahrer@mpia.de},
Charlotte Fairman$^{\orcidlink{0000-0001-9665-5260},2}$, 
James Kirk$^{\orcidlink{0000-0002-4207-6615},3}$,
Hannah R. Wakeford$^{\orcidlink{0000-0003-4328-3867},2}$,
Joanna K. Barstow$^{4}$, \newauthor
Anna B. T. Penzlin$^{5,3}$,
Lili Alderson$^{\orcidlink{0000-0001-8703-7751},6}$,
Richard A. Booth$^{\orcidlink{0000-0002-0364-937X},7}$, 
Duncan A. Christie$^{1}$,\newauthor 
Alastair B. Claringbold$^{8,9}$, 
Emma Esparza-Borges$^{\orcidlink{0000-0002-2341-3233},10,11}$,
Carlos Gasc\'on$^{\orcidlink{0000-0001-5097-9251}, 12}$,
Mercedes L\'opez-Morales$^{\orcidlink{0000-0003-3204-8183}, 13}$,\newauthor 
N. J. Mayne$^{14}$,
Mason McCormack$^{\orcidlink{0000-0002-1463-9847},15}$,
Annabella Meech$^{\orcidlink{0000-0002-7500-7173},12}$,
Paul Molli\`ere$^{\orcidlink{0000-0003-4096-7067},1}$,
James E. Owen$^{\orcidlink{0000-0002-4856-7837},3,16}$,\newauthor 
Vatsal Panwar$^{\orcidlink{0000-0002-2513-4465},8,9}$,
Denis E. Sergeev$^{\orcidlink{0000-0001-8832-5288},2}$,  
Daniel Valentine$^{\orcidlink{0000-0002-2643-6836},2}$,
Peter J. Wheatley$^{\orcidlink{0000-0003-1452-2240},8,9}$,
Maria Zamyatina$^{\orcidlink{0000-0002-9705-0535},14}$
\\
$^{1}$Max Planck Institute for Astronomy (MPIA), K\"{o}nigstuhl 17, 69117 Heidelberg, Germany \\
$^{2}$School of Physics, HH Wills Physics Laboratory, University of Bristol, Tyndall Avenue, Bristol BS8 1TL, UK\\
$^{3}$Department of Physics, Imperial College London, London SW7 2AZ, UK\\
$^{4}$The Open University, School of Physical Sciences, Milton Keynes MK7 6AA, UK\\
$^{5}$Ludwig-Maximilians-Universit{\"a}t M{\"u}nchen, Universit{\"a}ts-Sternwarte, Scheinerstr.~1, 81679 M{\"u}nchen, Germany\\
$^{6}$Department of Astronomy, Cornell University, 122 Sciences Drive, Ithaca, NY 14853, USA \\
$^{7}$School of Physics and Astronomy, University of Leeds, Leeds LS2 9JT, UK\\
$^{8}$Centre for Exoplanets and Habitability, University of Warwick, Gibbet Hill Road, Coventry CV4 7AL, UK\\
$^{9}$Department of Physics, University of Warwick, Gibbet Hill Road, Coventry CV4 7AL, UK\\
$^{10}$Instituto de Astrof\'isica de Canarias, C. V\'ia L\'actea, San Crist\'obal de La Laguna 38205, Spain\\
$^{11}$Department of Astrophysics, University of La Laguna, San Crist\'obal de La Laguna 38200, Spain\\
$^{12}$Center for Astrophysics | Harvard \& Smithsonian, 60 Garden St, Cambridge, MA 02138, USA\\
$^{13}$Space Telescope Science Institute, 3700 San Martin Drive, Baltimore, MD, USA\\
$^{14}$Department of Physics and Astronomy, Faculty of Environment, Science and Economy, University of Exeter, Exeter EX4 4QL, UK\\
$^{15}$Department of Astronomy \& Astrophysics, University of Chicago, Chicago, IL 60637, USA\\
$^{16}$Department of Earth, Planetary, and Space Sciences, University of California, Los Angeles, CA 90095, USA\\
}

\date{Accepted XXX. Received YYY; in original form ZZZ}

\pubyear{\the\year{}}

\begin{document}
\label{firstpage}
\pagerange{\pageref{firstpage}--\pageref{lastpage}}
\maketitle

\begin{abstract}
Hot Jupiters and their atmospheres are prime targets for transmission spectroscopy due to their extended atmospheres and the corresponding large signal-to-noise, providing the best possible constraints for the atmospheric carbon-to-oxygen (C/O) ratio and metallicity of exoplanets. 
Within BOWIE-ALIGN, we aim to compare JWST spectra of a sample of orbitally aligned and misaligned hot Jupiters orbiting F-type stars to probe the link between hot Jupiter atmospheres and planet formation history. 
Here, we present a near-infrared transmission spectrum of the aligned planet KELT-7b using one transit observed with JWST NIRSpec/G395H. We find weak features, only tentative evidence for \ce{H2O} and \ce{CO2} in the atmosphere of KELT-7b. This poses a challenge to constrain the atmospheric properties of KELT-7b and two possible scenarios emerge from equilibrium chemistry and free chemistry retrievals: a high-altitude cloud deck muting all features or an extremely low metallicity atmosphere, respectively. 
The retrieved C/O ratios from our data reductions range from $0.43 - 0.74$, while the atmospheric metallicity is suggested to be solar to super-solar ($1-16 \times$ solar). Although these wide constraints prevent detailed conclusions about KELT-7b's formation history, a solar-to-super-solar metallicity would imply the accretion of solid material during its formation, which is valuable information for the survey's wider goals of understanding the relative importance of gaseous to solid accretion.

\end{abstract}




\section{Introduction}
One of the strongest motivations behind studying the atmospheres of hot Jupiters is the possibility of linking their present-day atmospheres to their formation and evolution.  This led to the hypothesis that the formation location can be determined by the hot Jupiter's carbon-to-oxygen (C/O) ratio based on the ice lines in a protoplanetary disc, mostly motivated by \citet{Oberg2011TheAtmospheres}. However, many mechanisms and processes have since been suggested that can also alter the atmospheric C/O ratio of a hot Jupiter beyond the equilibrium chemistry approach. This includes planetary migration \citep[e.g.][]{Madhusudhan2014TowardsMigration, Booth2017ChemicalDrift, Penzlin2024BOWIE-ALIGN:Compositions}, disk ice lines evolving \citep[e.g.][]{Morbidelli2016,Owen2020}, and solids drifting within the disk \citep{Booth2017ChemicalDrift,Schneider2021HowC/O}. This is further complicated by the fact that models do not cover all the diversity found from observations of discs \citep[][]{Law2021}, and the unknown relation between solid and gaseous accretion \citep[e.g.][]{Espinoza2017}. 

With this high number of unknown and unconstrained parameters, it remains challenging to use C/O ratios as tracers for planet formation locations \citep[e.g.,][]{Molliere2022InterpretingAssumptions}. However, \citet{Penzlin2024BOWIE-ALIGN:Compositions} demonstrated that the opportunity lies in numbers: that is, by observing a statistical sample of two populations that in theory should have formed or migrated differently resulting in different C/O ratios, we could constrain planet formation models. While models cannot predict quantitive values, they do robustly predict qualitative relative trends between populations; together with \citet{Kirk2024BOWIE-ALIGN:History}, \citet{Penzlin2024BOWIE-ALIGN:Compositions} showed that C/O ratios and metallicities of hot Jupiters that have undergone disc migration should diverge from those that migrate after disc dispersal (disc-free or high-eccentricity migration) as disc-migrated planets accrete inner disc material  \citep[e.g., see also][]{Madhusudhan2014TowardsMigration, Booth2017ChemicalDrift}. 

Reliably and accurately measuring C/O ratios and metallicities in exoplanet atmospheres has proven challenging. For example, even when we can resolve both carbon- and oxygen-bearing species with JWST, we find that the retrieved values are influenced by retrieval setup \citep[e.g.,][, Welbanks et al., in prep.]{Lueber2024InformationWASP-39b} and detector offsets \citep[e.g.,][]{Ahrer2025TracingNIRSpec/G395H}. In addition, the limited range of atmospheric pressures probed with transmission spectroscopy might not be representative of the overall planetary composition \citep{Dobbs-Dixon2017}, and processes such as local atmospheric mixing \citep[e.g.,][]{Zamyatina24_quenchingdriven}, cloud formation \citep[e.g.,][]{Helling2016}, or a planet's interior evolution \citep{Muller2024} might further alter the atmospheric composition we are observing. 
Therefore, further systematic observational studies are required to address these theoretical and observational obstacles.

Following these studies and challenges, our survey `BOWIE-ALIGN'(JWST program ID: GO 3838, PIs: Kirk \& Ahrer) is set out to observe a sample of hot Jupiters, where half are in an aligned orbit, i.e., their orbital planes are perpendicular to the stellar axes meaning that they are aligned with the stellar equator; the other half of the sample is in a misaligned orbit around their host stars, i.e., exhibiting a significantly tilted orbital plane ($> 45^\circ$ by our definition). The aligned hot Jupiters are presumed to have migrated through the disc, while the misaligned ones are expected to have undergone high-eccentricity migration after the disc has dispersed. We determined our sample based on their orbital (mis)alignment in addition to only considering hot Jupiters that orbit F stars above the Kraft break (effective temperatures $\gtrsim 6100$\,K). These stars have radiative outer envelopes that result in inefficient tidal realignment \citep[e.g.,][]{Albrecht2012ObliquitiesMisalignments}, which reduces the possibility of our aligned sample falsely including initially misaligned planets (via high-eccentricity migration) that have had their obliquities damped. 

In this work, we present the JWST NIRSpec/G395H observations of \myplanet, the third planet in our programme. For first planet, WASP-15b (misaligned), \citet{Kirk2025BOWIE-ALIGN:WASP-15b} found that its atmospheric metallicity is super-solar while its C/O ratio is consistent with solar, implying planetesimal accretion. \citet{Kirk2025BOWIE-ALIGN:WASP-15b} further reported evidence for the photochemical product \ce{SO2} in the atmosphere of WASP-15b. \citet{Meech2025BOWIE-ALIGN:Spectroscopy} found that the observations of the second planet, TrES-4b (aligned), suggest that its atmospheric C/O ratio and metallicity are subsolar. This points towards oxygen-rich gas accretion or to a combination of carbon-poor solid and low-metallicity gas accretion.

\myplanet was discovered by \citet{Bieryla2015} and has a mass of $1.39 \pm 0.22$\,M$_\mathrm{Jup}$ and a radius of $1.60 \pm 0.06$\,R$_\mathrm{Jup}$ \citep{Stassun2017Parallaxes}. Its equilibrium temperature has been estimated as $2048 \pm 27$\,K \citep{Bieryla2015}. With these planetary parameters, \myplanet is on the boundary of being a hot or ultra-hot Jupiter as combined with its surface gravity its dayside may reach temperatures where most molecules dissociate \citep{Parmentier2018FromContext}. The parameters of its host star, \mystar, are summarised in Table\,\ref{tab:stellar-parameters}. Based on its measured obliquity of $-10.55 \pm 0.27^\circ$ \citep{Tabernero2022KELT7align}, \myplanet belongs to the aligned sample, i.e., the sample within the BOWIE-ALIGN survey to have migrated within their protoplanetary discs.

\begin{table}

    \centering
    \caption{Stellar and planetary parameters for the \mystar planetary system. References are as follows: [1] \citet{Bieryla2015}, [2,3] \citet{Cannon1918The3h,Cannon1993VizieR1989}, [4] \citet{Stassun2017Parallaxes}, [5] \citet[Gaia DR3,][]{GaiaCollaboration2023GaiaProperties}, [6] \citet[Gaia DR3 GSP-Phot,][]{Andrae2023GaiaPhotometry}, [7] \citet{Bieryla2015}, [8] \citet{Patel2022EmpiricalTESS}, [9] \citet{Tabernero2022KELT7align}. }
    \begin{tabular}{l c c }
    \hline
        Parameter & Value &   Ref.\\ \hline
        Spectral type & F/F2  & [1/2, 3]\\
        Effective Temperature, $T_{\textrm{eff}}$ (K) & $6768 \pm 7$ & [4]\\
        Age (Gyr) & $1.3 \pm 0.2$  & [1] \\
        Surface gravity, log $g$ (log$_{10}$(cm/s$^{2}$)) & $4.149 \pm 0.019$  & [1]\\
        Metallicity [Fe/H] (dex) & $0.139^{+0.075}_{-0.081}$ &  [1] \\
        Stellar Mass, M$_\textrm{*}$ (M$_\odot$) & $1.76 \pm 0.25$ & [4]\\
        Stellar Radius, R$_\textrm{*}$ (R$_\odot$) & $1.768 _{-0.029}^{+0.013}$ & [5,6] \\
        \hline
        Planetary Mass, M$_\mathrm{p}$ (M$_\mathrm{Jup}$) & $1.39 \pm 0.22$ & [4] \\   
        Planetary Radius, R$_\mathrm{p}$ (R$_\mathrm{Jup}$) & $1.60 \pm 0.06$ & [4] \\  
        Equ. Temperature, T$_\mathrm{eq}$ (K) & $2048 \pm 27$ & [7] \\   
        Orbital period, P (days) & $2.7347703^{+0.0000037}_{-0.0000039} $ & [8] \\
        Obliquity, $\lambda$ (degrees) & $-10.55 \pm 0.27^\circ$ & [9] \\
        \hline
    \end{tabular}
    
    \label{tab:stellar-parameters}
\end{table}

\myplanet's atmosphere has been studied previously. 
\citet{Pluriel2020ARES.WFC3} found evidence for \ce{H2O} in transmission and emission using the Hubble Space Telescope (HST) Wide Field Camera 3 (WFC3) G141 grism ($1.1-1.7$\textmu m). The same observations also suggested the presence of \ce{H-} absorption which was further confirmed by \citet{Changeat2022FiveEclipse}. Recently, \citet{Gascon2025TheKELT-7bb} presented a new HST WFC3/UVIS (ultraviolet-visible) G280 featureless transmission spectrum covering $0.19-0.8$\textmu m. In combination with the HST G141 data and the \ce{H-} absorption features, \citet{Gascon2025TheKELT-7bb} also show evidence for water dissociation in their analysis. Under the assumption of thermochemical equilibrium, hot Jupiters with temperatures $<2500$\,K may not exhibit strong \ce{H-} absorption features \citep{Kitzmann2018OpticalAtmospheres}, though a combination of photochemical and collisional processes may lead to a higher production of \ce{H-} at temperatures $<2500$\,K \citep{Lewis2020IntoRevealed}, especially on hot Jupiters orbiting F stars with intense UV irradiation. The evidence of dissociation of \ce{H2O} in \myplanet's atmosphere may affect any abundance constraints of \ce{H2O} retrieved using JWST observations and thus influence the inferred C/O ratio and metallicity. 

Other studies such as high-resolution atmospheric observations of \myplanet have not shown any significant detections of atomic species in the atmosphere. \citet{Tabernero2022KELT7align} used HORuS (the High Optical Resolution Spectrograph) mounted on the GTC (Gran Telescopio Canarias) to search for the presence of H$\alpha$, Li I, Na I, Mg I, and Ca II in \myplanet's atmosphere, but were only able to place upper limits on their abundances. Similarly, \citet{Sicilia2025TheExoplanets} did not detect Na I features using TNG/HARPS-N observations.

This manuscript is structured as follows. First, we discuss the observations in Section\,\ref{sec:observations}, which is followed by the description of the data reduction and analysis in Section\,\ref{sec:data_reduction}. Our atmospheric retrieval setups are laid out in Section\,\ref{sec:retrievals}, followed by the discussion of our results in Section\,\ref{sec:results} and our conclusions in Section\,\ref{sec:discussion-conclusions}.

\section{Observations}
\label{sec:observations}
Our JWST observations of \myplanet took place on 3 February 2024 using NIRSpec/G395H, NRSRAPID readout in the Bright Object Time Series (BOTS) mode. We used 5 groups/integrations, with 4976 total integrations and an overall observing time of 7.51 hours, consisting of 2.35 hours pre-transit, 3.51 hours in-transit and 1.66 hours post-transit time. The data was split into four segments.
After ingress, at around one third of the transit time (around BJD\_MJD 60363.874, integration \# 2365), we observed a jump in the data, suggesting that a mirror tilt event took place. We investigated this event further in Section\,\ref{sec:mirror-tilt-event}. This mirror tilt event took place in a period where there was a resurgence of these events observed and only a few weeks prior to the largest seen event seen in the JWST science era\footnote{\href{https://jwst-docs.stsci.edu/jwst-observatory-characteristics-and-performance/optics-performance-and-stability}{JWST User Documentation}; accessed 5 August 2025} \citep{Perrin2024Title:2}.

\section{Data reduction}
\label{sec:data_reduction}

To reduce our data, we used three independent pipelines: \eureka, \exoticjedi and \tiberius. This enabled us to check whether our resulting conclusions were independent of the method used to reduce the data and to fit the light curves with the tilt event. We describe the approaches used during the independent reductions in the following subsections, following standard reduction and analysis process routines with the exception of the additional fitting of the mirror tilt-event.

\subsection{\eureka}
\eureka is an open-source pipeline \citep{Bell2022Eureka}\footnote{We used version v0.11.dev245+ge8ea1d1c.d20240701.} for the reduction of time-series observations with JWST as well as HST. It has been benchmarked against other pipelines and has been successfully used in multiple JWST data sets \citep[e.g.][]{Ahrer2023, Bell2023MethaneWASP-80b,MoranStevenson2023, Beatty2024SulfurB, Kirk2025BOWIE-ALIGN:WASP-15b,Meech2025BOWIE-ALIGN:Spectroscopy}. 

\subsubsection{Light curve extraction}

We started our \eureka analysis with the \texttt{uncal.fits} files. \eureka Stages\,1 and 2 are wrapped around the default \texttt{jwst} pipeline (v1.12.2, context map 1242) steps. We followed the default steps (with a jump step threshold of $10\sigma$) in addition to 1/f background subtraction at the group level and we opted for using a custom scale factor (using a smoothing filter calculated from the first group) for the bias correction. 

In Stage\,3 of \eureka we extract the time-series 1D spectra. We masked outliers greater than $5$ times the median in the spatial direction and used double-iterative masking with $>5\sigma$ along the time axis. We correct for the curvature of the trace and subtract the background by fitting a constant for each frame to the area that is $>6$ pixels away from the central pixel of the spectral trace. Within the \texttt{jwst} pipeline each integration receives a calibrated wavelength map from which we extract the wavelength at the source position.
Finally, we perform optimal spectral extraction \citep{Horne1986AnSpectroscopy.} using a full width of 9 pixels. This is followed by \eureka's Stage\,4, where we bin our data to the respective wavelength resolutions (R=100 and R=400) and clip outliers $>5\sigma$ based on a box car filter width of 20\,pixels used to calculate a rolling median. We also generated a white light curves for NRS1 and NRS2.

\subsubsection{Mirror tilt event}
\label{sec:mirror-tilt-event}

During our JWST observations of \myplanet, a mirror tilt event occurred, causing a change in the flux in our light curves. A tilt event is defined as an abrupt change in the position of a mirror segment that happens occasionally at varying magnitudes \citep[e.g., see][]{Rigby2023TheCommissioningb}. This was also seen in previous transit observations, e.g., in HAT-P-14b \citep{Schlawin2023JWSTTilts} and WASP-39b \citep{Alderson2023EarlyG395H}. 
\citet{Alderson2023EarlyG395H} find that either fitting a simple step model or detrending the light curve using trace properties such as Full Width Half Maximum (FWHM) or trace position worked equally well. In our \exoticjedi light curve fitting we test detrending against x and y position and while it accounted for some variations due to the tilt event, we required an additional step function to model the light curve adequately. We note that in contrast to the WASP-39b tilt event the effect was not clearly visible in the FWHM or positional tracers of our \myplanet observations. This is likely because the tilt event observed here is much smaller in magnitude and the data exhibits more noise, likely due to the smaller number of groups. 

\begin{figure}
    \centering
    \includegraphics[width=\linewidth]{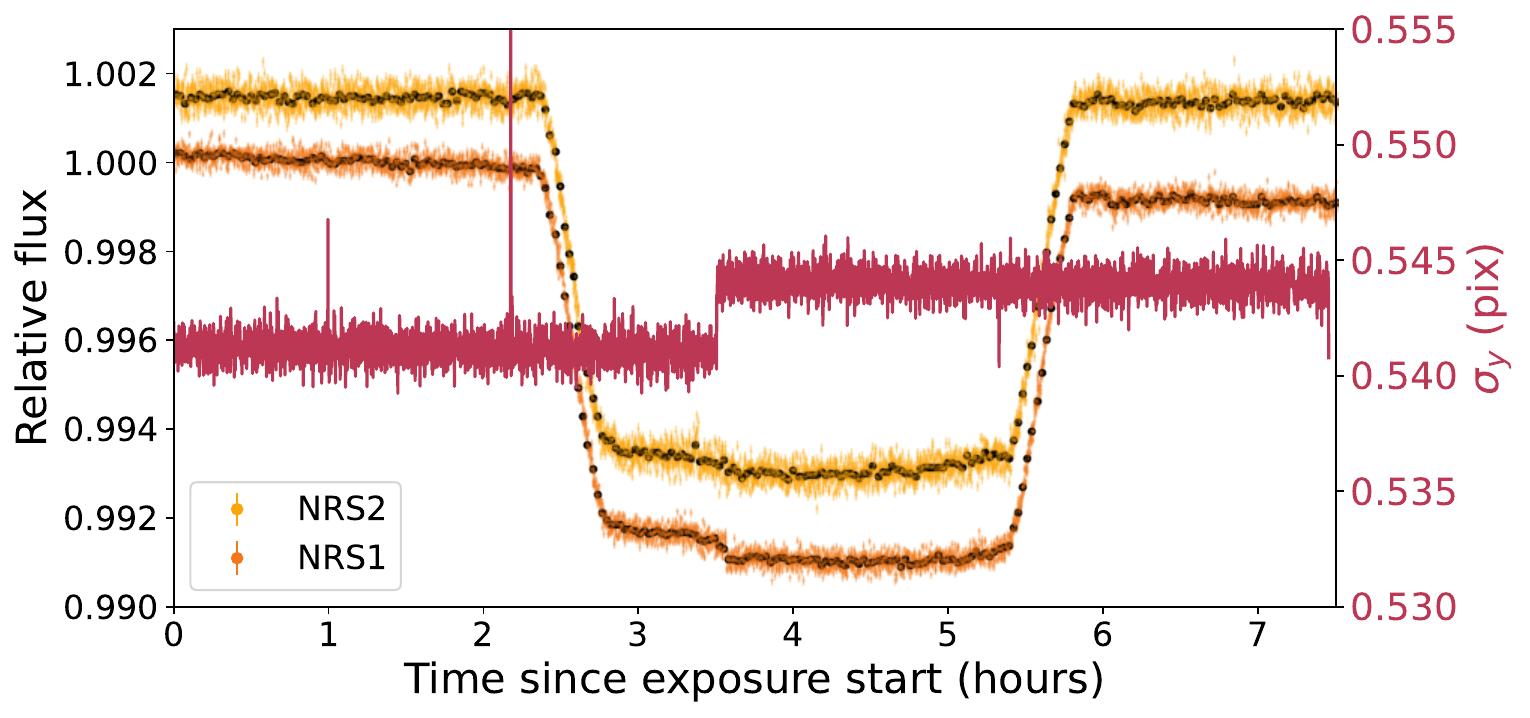}
    \caption{\myplanet transit white light curve of NRS2 (top, light orange) and NRS1 (bottom, dark orange) compared to one of the fitted widths ($\sigma_x$, dark magenta) to the JWST guidance data using the python package \texttt{spelunker} \citep{DealEspinoza2024spelunker}, binned to the time stamps of our light curve. We find that there is a definite tilt event visible in both. However, the time stamp of the event occurring is offset at $\sim5$\,minutes between the guidance data and our light curve. We attribute this to calibration differences between FGS and science data. }
    \label{fig:spelunker_tilt-event}
\end{figure}

We investigated this event by looking at the JWST NIRISS Fine Guidance Sensor (FGS) data, which provides guide star imaging data in parallel to every JWST observation. For this purpose, we use the python package \texttt{spelunker} \citep{DealEspinoza2024spelunker}, developed to enable access and provide useful tools to analyse these guide star data products. We fit a 2D Gaussian to extract the guide star flux and bin it to the time stamps of our transit. The time sampling of the FGS is at 64\,ms per exposure, while the time between our integrations is $5.43$\,s. Then we compare the fitted width of the guide star ($\sigma_x$) to the NRS1 and NRS2 light curves, see Fig.~\ref{fig:spelunker_tilt-event}. We find a definite jump in the guide star width, however, it does not exactly line up with the time of the event seen in the light curve. It is $\sim5$ minutes apart, which we attribute to the fact that the FGS time stamps are not calibrated in the same way as the scientific observations are and can show offsets of that order (confirmed by private communication with N. Espinoza (STScI)). This is further confirmed as the last time stamp from the FGS data is $3$\,min before the end of our observations (see also Fig.\,\ref{fig:spelunker_tilt-event}). Thus, while we try using this data for the detrending of the light curves at a later stage (with a time offset), it resulted in the same result as simply fitting a step function so we elected to use the latter, simpler method.

\begin{figure*}
    \centering
    \includegraphics[width=\linewidth]{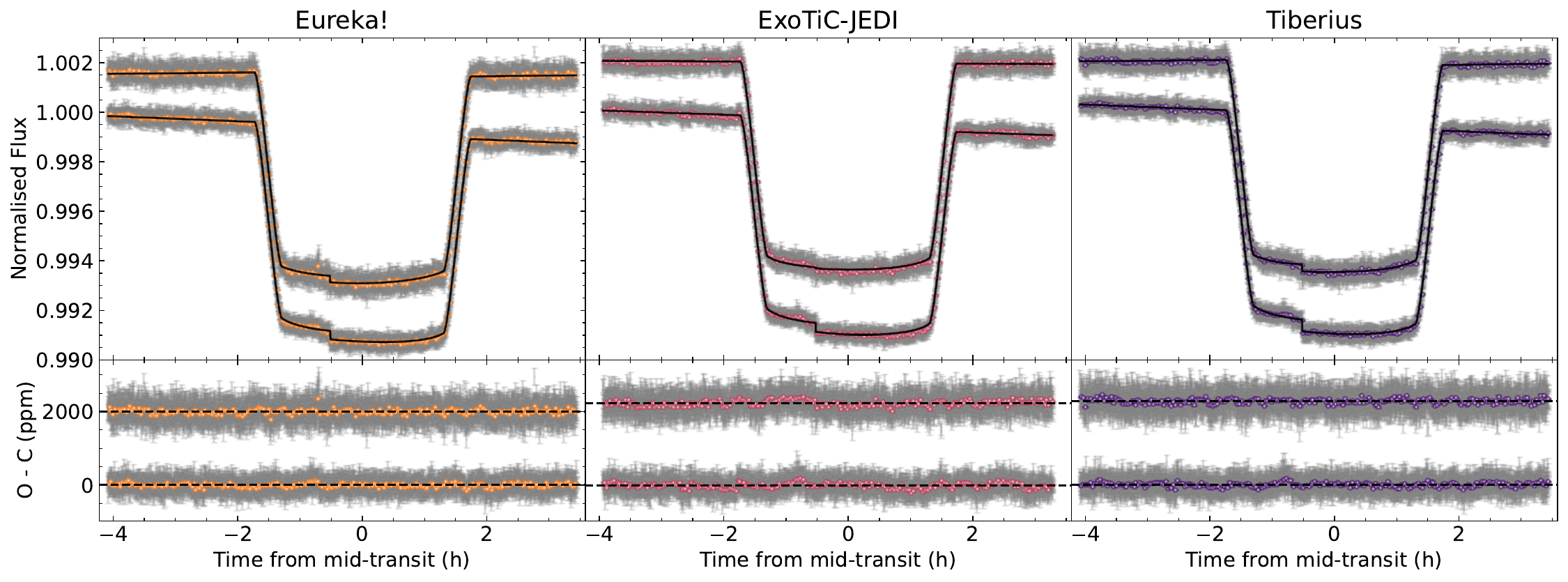}
    \caption{Top: Transit broadband light curves of \myplanet with NIRSpec/G395H using \eureka, \exoticjedi and \tiberius (from left to right), with NRS2 being the top (offset by 0.002) and NRS1 the bottom light curve. Bottom: The corresponding residuals using the best-fit model; the top data showing the NRS2 residuals (offset by 2000\,ppm) and NRS1 residuals on the bottom. }
    \label{fig:wl-curves}
\end{figure*}

\subsubsection{Light curve fitting}

We fit our extracted light curves using \eureka's Stage\,5 using a \texttt{batman} transit light curve model \citep{batman} and a linear in time systematic model. In addition, we include a simple step function to account for the mirror tilt event. We explore the parameter space using the MCMC sampling algorithm \texttt{emcee} \citep{Foreman-Mackey2013EmceeHammer} using 300 walkers, 1000 steps and an additional 600 burn-in steps, where the starting parameter values are set as the resulting values from an initial least-squares fit. 

First, we fit the NRS1 (2.87 - 3.72 \textmu m) and NRS2 (3.82 - 5.16 \textmu m) white-light light curves independently and retrieve the system parameters scaled stellar radius (a/R$_*$), inclination and mid-transit time as well as R$_\mathrm{p}$/R$_*$. We fixed the period to 2.7347703\,days \citep{Patel2022EmpiricalTESS} and assumed a circular orbit (e=0, $\omega = 90^\circ$). We further fit for the step time and step amplitude to account for the mirror tilt. The retrieved system parameters and the step time for NRS1 and NRS2 were then held fixed to those values when fitting the spectroscopic light curves. These retrieved system parameters and their corresponding uncertainties are given in Table\,\ref{tab:system_params}.

\begin{figure*}
    \centering
    \includegraphics[width=\linewidth]{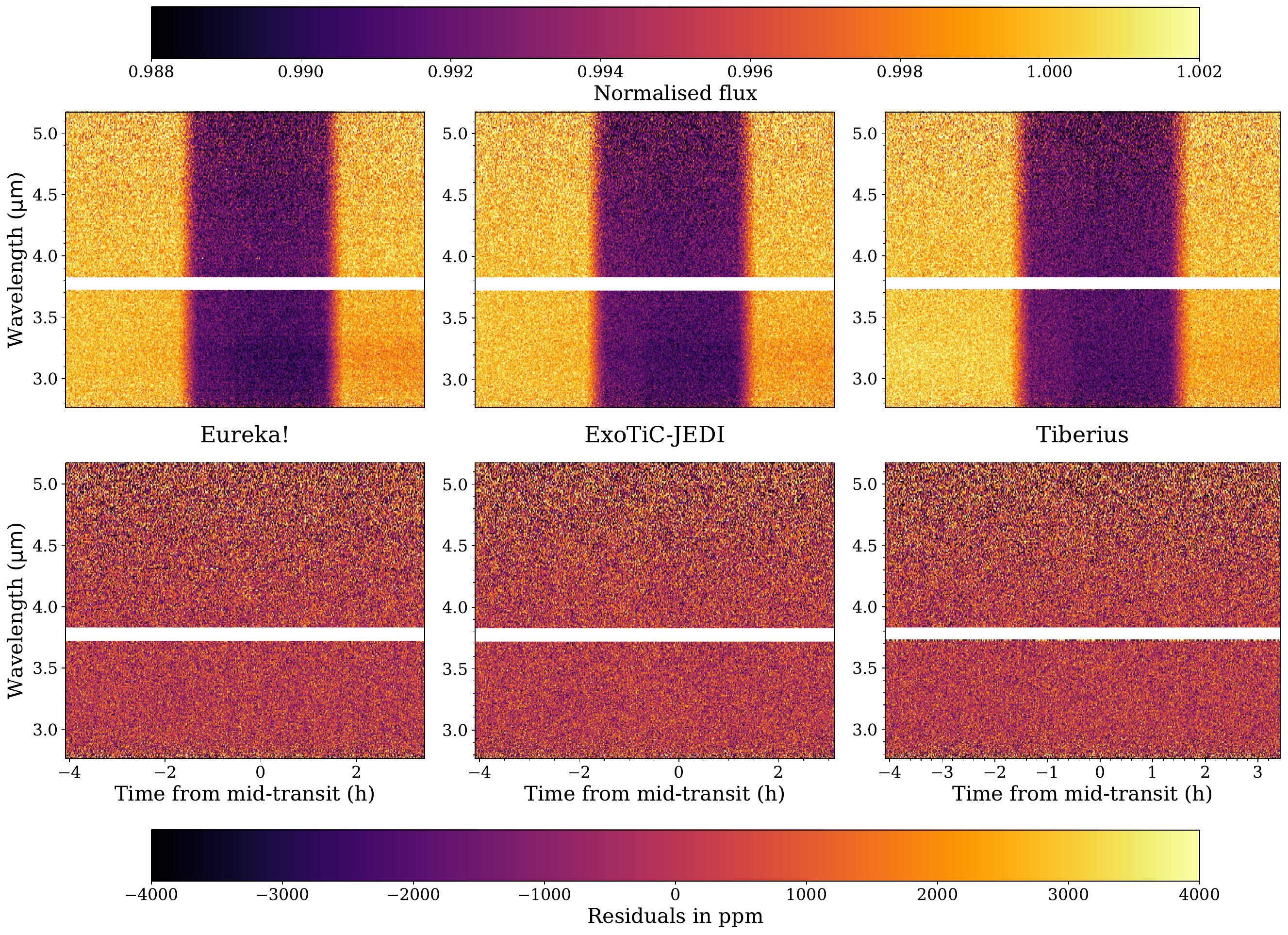}
    \caption{Extracted raw spectroscopic light curves (top row) and corresponding residuals (bottom row) of each reduction (columns; \eureka, \exoticjedi and \tiberius) of \myplanet with NIRSpec/G395H at a resolution of R=400. The white area corresponds to the gap between the two detectors (NRS1 and NRS2).}
    \label{fig:light-curves-2d}
\end{figure*}

For the modelling of the limb-darkening, we use the quadratic limb-darkening law and 
use both limb-darkening parameters (u1, u2) as free fitting parameters in all light curve fits, with a broad uniform prior. The fitted white light curves are shown in Fig.\,\ref{fig:wl-curves}, while the spectroscopic light curves at R=400 are shown in Fig.\,\ref{fig:light-curves-2d} together with the respective residuals. 

We further conduct a combined fit of the NRS1 and NRS2 white light curves to investigate system parameters and their differences in NRS1 and NRS2. In this fit we again model the light curve using quadratic limb-darkening (freely fit for u1 and u2), a linear model and the transit model. The fitted parameters are also shown in Table\,\ref{tab:system_params}.

\begin{table*}
    \caption{System parameters for the transit of \myplanet from the JWST NIRSpec/G395H white light curves as fitted by the three independent reductions.   }
    \label{tab:system_params}
    \centering
    \begin{tabular}{lccccc} \hline
         Pipeline & Detector & $T_0$ (BJD) & $R_\mathrm{p}/R_*$ & $a/R_*$ & $i$ (\degree) \\ \hline 
         \eureka & NRS1 & $2460364.395942 \pm 0.000016$ & 	$0.090375 \pm 0.000053$ & $5.5528 \pm 0.0030$ & $83.833^{+0.0091}_{-0.0093}$ \\ 
         \eureka & NRS2 & $2460364.395921 \pm 0.000022$ & 	$0.090236 \pm 0.000058$ &  $5.5596 \pm 0.0033$ & $83.847^{+0.0093}_{-0.0099}$ \\ \hline
         \exoticjedi & NRS1 & $2460364.396044 \pm 0.000016$ & $0.090069 \pm 0.000032$ & $5.551 \pm 0.011$ & $83.83 \pm 0.03$ \\ 
         \exoticjedi & NRS2 & $2460364.396051 \pm 0.000020$ & $0.090283 \pm 0.000039$ & $5.565 \pm 0.013$ & $83.85 \pm 0.04$  \\ \hline
         \tiberius & NRS1 & $ 2460364.395937 \pm 0.000015 $ & $ 0.090487 \pm 0.000030 $  & $ 5.5014 \pm 0.0097  $  & $ 83.673 \pm 0.031 $ \\ 
         \tiberius & NRS2 & $ 2460364.395944 \pm 0.000020  $ & $ 0.090294 \pm 0.000038 $ & $ 5.5471^{+0.0124}_{-0.0121} $ & $ 83.804 \pm 0.040 $ \\ \hline
         \eureka combined fit & NRS1 \& NRS2 & $2460364.395933 \pm 0.000013$ & & $5.5538 \pm 0.0030$ & $83.8317 \pm 0.0092$ \\ 
          & NRS1 & & $0.090389 \pm 0.000051$ \\ 
          & NRS2 & & $0.090242 \pm 0.000059$ \\ \hline
         \multicolumn{2}{l}{\citet{Patel2022EmpiricalTESS}} & & & $5.60 \pm 0.07$ & $83.92 \pm 0.19$  \\ \hline
    \end{tabular}
\end{table*}

\subsection{\exoticjedi}
\exoticjedi \citep{Alderson2022Exo-TiC/ExoTiC-JEDI:V0.1-beta-release} is an end-to-end data analysis toolkit to go from \texttt{uncal} spectroscopic files to processed planetary spectra and has been implemented in a number of exoplanet studies (e.g., \citetalias{MayMacDonald2023ApJ}; \citealt{Alderson2023EarlyG395H,Alderson2024JWSTTOI-836b,Scarsdale2024AJ}). Stages 1 and 2 act as a custom wrapper for the official \texttt{jwst} pipeline (v1.13.4, context map 1364) with custom steps incorporated for bias correction, and cleaning of 1/f noise at the group level. We set the jump step level to 15-sigma which has been found to balance the number of detections without over-correcting for changes in flux at the group level \citep[e.g.,][]{Alderson2023EarlyG395H}. In Stage 3, \exoticjedi performs a series of cleaning steps. We use the data quality (DQ) flags from the \texttt{jwst} pipeline that indicate ``do\_not\_use", ``dead", ``hot", ``saturated", ``low\_qe", and ``no\_gain\_value'' to identify bad pixels and replace them with an average of the neighbouring pixels, this results in 0.3\% and 0.5\% of pixels being replaced in NRS1 and NRS2 respectively. We then perform custom routines to remove outliers in time and spatial axis, where corrections are made using the adjacent 4 pixels either side of the effected pixel in the row (for spatial) and in the integration (for time) to the flagged pixel. For the time outliers, we use a window of ten frames and a 20-sigma threshold to identify outliers and for spatial variations, we use a 6-sigma cut off in each integration evaluated with a 21 pixel window. We additionally correct for 1/f noise at the integration level by masking the spectral trace and subtracting the median value on a column-by-column basis. We extract the final stellar spectrum using an aperture of five times the full-width-half-maximum of the PSF, which is approximately 0.7 pixels wide, resulting in a total aperture width of $\sim$7 pixels, and use intrapixel extraction to account for partial flux illumination on the edge pixels. 


\subsubsection{Light curve fitting}
Using \exoticjedi we fit the light curves from NRS1 and NRS2 separately throughout the process. We first fit for our broadband integrated light curves which span 2.87--3.72\,$\upmu$m for NRS1 and 3.82--5.12\,$\upmu$m for NRS2. In each broadband light curve, we fit for the centre of transit time, transit depth, a/R*, and inclination (see Table\,\ref{tab:system_params}), while keeping orbital period (2.7347703\,days), eccentricity (0.0), and omega (90) fixed. In order to correct for the mirror tilt-event, we fit for the time of the flux jump and the amplitude of the change in flux, correcting with a simple step function. We additionally fit for standard systematics from the detector and telescope which has the following functional form, 
$$S(t) = s_0 + (s_1 \times (\mathrm{xpos} \times |\mathrm{ypos}|)) + (s_2 \times t) \mathrm{,}$$
where $s_0$, $s_1$, and $s_2$ are variables to be fit for, xpos and ypos are the changing position of the spectrum on the detector calculated by cross-correlating the spectra to a template (note we take the absolute value of the y-position in this model), and $t$ is the time array. This systematic model along with the mirror tilt step-function is then multiplied with a \texttt{batman} light curve model using a least-squares minimizer to determine the best-fit parameters. For our spectroscopic light curves we then fix the a/R$_*$, inclination and center of transit time to those determined from the broadband fits and fit for the transit depth in each bin. 
For each light curve fit, we fix the 4-parameter non-linear limb-darkening coefficients to those computed using \texttt{ExoTiC-LD} \citep{Grant2024ExoTiC-LD:Coefficients} using the stellar parameters shown in Table\,\ref{tab:stellar-parameters} to interpolate the MPS1 stellar model grid \citep{Kostogryz2023RNAAS}. The spectroscopic light curves at R=400 and respective residuals are shown along the \eureka and \tiberius ones in Fig.\,\ref{fig:light-curves-2d}.
We compute the transmission spectrum for R=100 and R=400 allowing the systematic model and mirror tilt step-function to be freely fit in each bin. Across all of our spectroscopic bins we find that there is negligible correlated (red) noise and that our residuals bin down with the expected photon precision.

\subsection{\tiberius}

We used the \texttt{Tiberius} \citep{Kirk2017,Kirk2021} JWST data reduction software for a third reduction, which has been used in a number of JWST studies \citep[e.g.,][]{Alderson2023EarlyG395H,Lustig-Yaeger2023,Kirk2024JWST/NIRCam341b}. As described in \cite{Kirk2024BOWIE-ALIGN:History}, we are performing one identical reduction for each BOWIE-ALIGN target to help avoid reduction-dependent biases when we combine the spectra at the end of this survey, which is this \texttt{Tiberius} reduction.

\subsubsection{Light curve extraction}

 For this reason, our \texttt{Tiberius} (v1.0.4) reduction used the same extraction input file, \texttt{jwst} pipeline version (v1.8.2), JWST calibration files and light curve binning schemes as our \texttt{Tiberius} extractions of WASP-15b \citep{Kirk2025BOWIE-ALIGN:WASP-15b} and TrES-4b \citep{Meech2025BOWIE-ALIGN:Spectroscopy}. We refer the reader to \cite{Kirk2025BOWIE-ALIGN:WASP-15b} for a detailed explanation of the spectral extraction process. In brief, we perform our own 1/f correction at the group-level stage followed by custom bad pixel and cosmic ray correction routines. Finally, we perform spectral tracing using a fourth order polynomial and standard aperture photometry with an aperture full width of 8 pixels at the integration stage.

\subsubsection{Light curve fitting}

For the reasons described above, we initially followed the same light curve fitting procedure as used in previous BOWIE-ALIGN analyses. This involved fitting the white light curves with a combination of a \texttt{batman} analytic transit light curve \citep{Kreidberg2015BatmanPython} and a linear-in-time systematics model. However, as explained in Section \ref{sec:mirror-tilt-event}, there is a mirror-tilt event in the transit that was not detrended by the adoption of a simple linear polynomial. For this reason, we added a step function to our systematics model, parametrised by the step position (in units of integration number) and a normalising term which allows the systematics model to move up and down in flux after the step. 

The free parameters of the transit model are the time of mid-transit ($T_0$), inclination of the planet ($i$), the ratio of semi-major axis to stellar radius ($a/R_*$) and ratio of planet-to-star radii ($R_P/R_\star$). Like our \eureka reduction, we assume quadratic limb darkening however, unlike our \eureka reduction, we fix the coefficients to the values computed with \texttt{ExoTiC-LD} \citep{Grant2024ExoTiC-LD:Coefficients} and 3D stellar models \citep{Magic2015} using the stellar parameters in Table \ref{tab:stellar-parameters}. The free parameters of the systematics model were the two coefficients of the linear polynomial and the step function's breakpoint and normalising term. Therefore, our white light curve fits had a total of 8 free parameters. We held the planet's eccentricity fixed to 0 and the longitude of periastron to 90$^\circ$ \citep{Bieryla2015}. The period was fixed to 2.7347703\,d \citep{Patel2022EmpiricalTESS}.

We found that Levenberg-Marquadt was unable to reliably estimate the uncertainties in the step functions' parameters. For this reason, we used MCMC sampling through \texttt{emcee} \citep{Foreman-Mackey2013EmceeHammer}, unlike in our previous BOWIE-ALIGN analyses where we used Levenberg-Marquadt. We ran two sets of chains, both with 180 walkers and ran for as many steps as needed to reach $>50\tau$ where $\tau$ is the autocorrelation time. Typically, 4000 steps for each set of chains was sufficient to reach this criterion. The best fit from the first iteration was used to rescale the photometric uncertainties to give $\chi^2_\nu = 1$ for the best-fit model. The second iteration was randomly scattered around the median values from the first iteration. We then took the medians, 16th and 84th percentiles from the final 15000 steps of the second iteration, with a thinning factor of 10, as our parameter values and uncertainties. 

The system parameters we obtain are reported in Table \ref{tab:system_params}. As this Table shows, we found a disagreement between the system parameters derived from the NRS1 and NRS2 detectors, particularly $a/R_\star$ and $i$. This discrepancy between the two detectors' system parameters is not seen in either the \eureka or \exoticjedi reductions. For our other BOWIE-ALIGN targets (WASP-15b and TrES-4b), we did not see this inconsistency between the detectors. In an attempt to bring these into better agreement, we experimented with 25 variations of our model, including additional systematics parameters such as $x$ and $y$ position, fine guide star data, free limb darkening, high order polynomials, clipping data around the step function, and choice of starting positions for the chains. None of these tests brought the NRS1/NRS2 system parameters into agreement. For this reason, we proceeded with our original model since it is as similar as possible to our other BOWIE-ALIGN analyses. 

We adopted the same model and MCMC sampling procedure for our spectroscopic light curve fits. Normally at this stage in our uniform \tiberius analyses of BOWIE-ALIGN targets, we would fix the system parameters ($a/R_*$, $i$ and $T_0$) to the weighted mean white light values from NRS1 and NRS2. However, as discussed, these were inconsistent between NRS1 and NRS2. For this reason, we fixed the system parameters in our spectroscopic light curve fits to the values we infer from our NRS2 white light curve fit. We chose these values due to the better agreement with the system parameters derived from \eureka and \exoticjedi which suggests the problem with the \tiberius parameters arises from NRS1, which could be related to the larger amplitude red noise for this detector and dataset. At this stage, we also fixed the step function's break point, $s_p$, to the value from the \texttt{Tiberius} NRS2 white light curve fit. This gave us four free parameters per spectroscopic light curve, $R_P/R_{*,i}$, the two coefficients of the linear polynomial and the step function normalising term. The light curves for the wavelength binning of R=400 and the residuals following the transit model fitting are shown in the right panel of Fig.\,\ref{fig:light-curves-2d}.

\subsection{Data reduction comparisons}

\begin{figure*}
    \centering
    \includegraphics[width=\linewidth]{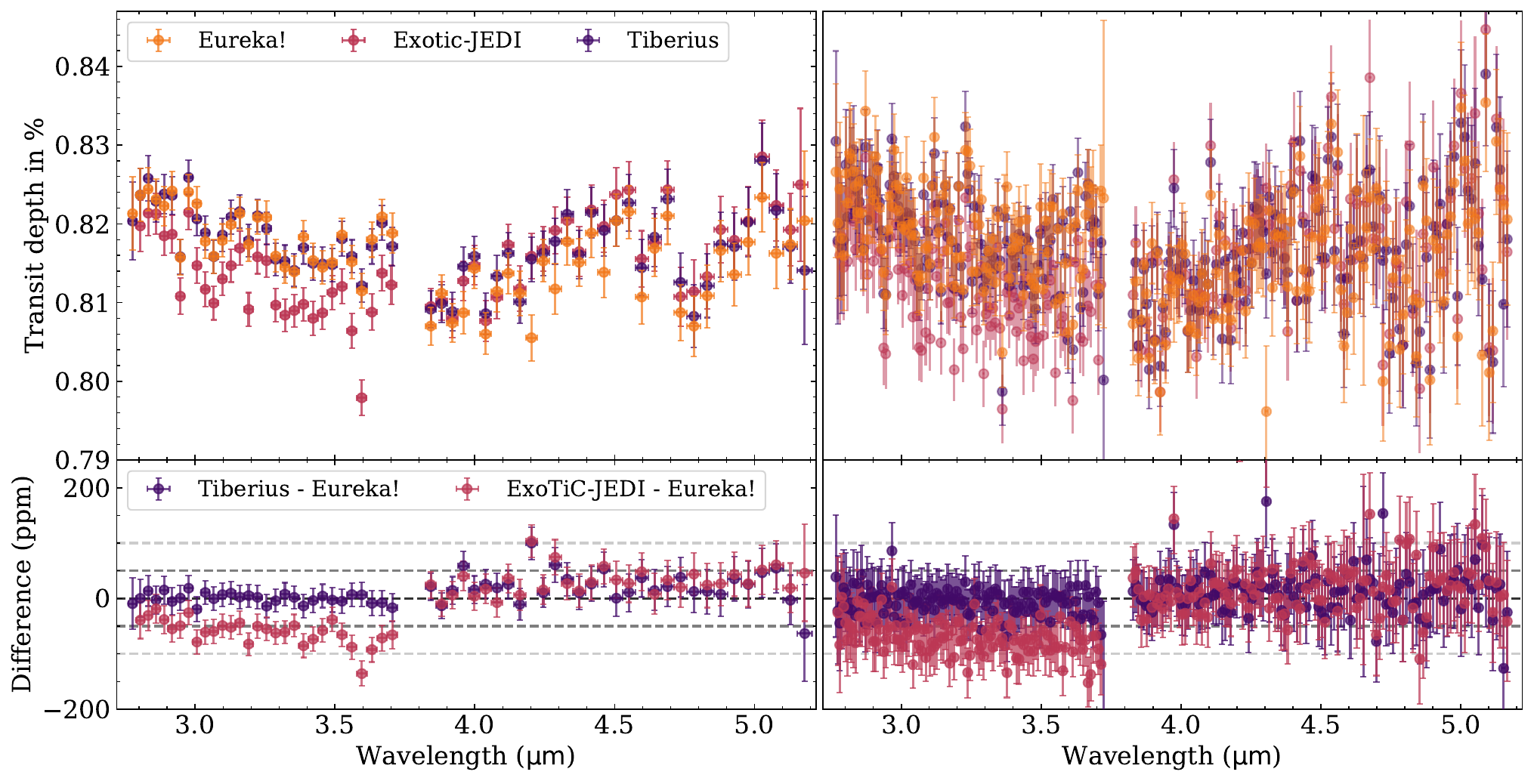}
    \caption{Top: \myplanet's transmission spectrum using JWST's NIRSpec/G395H and three independent reductions \eureka, \tiberius and \exoticjedi, in two resolutions R=100 (left) and R=400 (right). Bottom: We find offsets between the reductions likely due to the small number of groups/integrations and mirror tilt event. The differences are shown here relative to the \eureka reduction and horizontal lines are indicated at 0, $\pm 50$, $\pm100$ ppm, which are typical ranges of offsets found by retrievals. } 
    \label{fig:transmission-spectrum_all}
\end{figure*}

Fig.\,\ref{fig:transmission-spectrum_all} shows the transmission spectra of \myplanet as determined by the three independent pipelines at resolutions R=100 and R=400. Note that we find offsets between the reductions and between NRS1 and NRS2, which we attribute to a combination of the low number of groups/integrations, the mirror tilt event and differences in the treatment of limb-darkening. On the other hand, when looking at the fitted NRS1 and NRS2 white light curves in Fig.\,\ref{fig:wl-curves} we find that the fitted amplitude of the step caused by the tilt event in \exoticjedi's NRS2 is much smaller compared to \eureka's and \tiberius'. Considering the differences between \exoticjedi and the other reductions is smaller in NRS2 than NRS1, we can likely rule out light curve fitting as the cause for the offset. 

In terms of quantitative comparisons, all three reductions showed similar transit depth precisions and scatter. The median transit depth uncertainties are 53, 50 and 51 ppm and the averages of the root mean square of the residuals are 1601, 1578 and 1613 ppm for \eureka, \exoticjedi and \tiberius, respectively, calculated across both NRS1 and NRS2.

The fitted system parameters did not indicate any reason for offsets between the reductions. We further tested this by retrieving a transmission spectrum using the system parameters from the combined NRS1\&NRS2 white light fit using \eureka, but this did not amount to significant offset differences, see Appendix\,\ref{sec:appendix_WL_combined}.

Thus, we choose to use all transmission spectra of all reductions and resolutions to investigate inferences of \myplanet's atmosphere, but we set our fiducial spectrum as \eureka R=400.

\section{Atmospheric Retrieval Analysis}
\label{sec:retrievals}

\subsection{Retrieval Setup}
We use three different atmospheric retrieval codes to interpret our JWST NIRSpec/G395H transmission spectrum of \myplanet. We use \texttt{petitRADTRANS} (\pRT) and \poseidon to run equilibrium chemistry retrievals and free chemistry for retrievals on the JWST data, on all reductions and resolutions, while we use the third framework, \nemesispy, to model the panchromatic JWST+HST spectrum of \myplanet. For all retrieval setups, we fix stellar parameters to the values presented in Table\,\ref{tab:stellar-parameters} and planetary parameters from Table\,\ref{tab:system_params}.  

Our free chemistry setup is further split into two cases: (1) only including near-infrared, \textit{base} species which are \ch{H2O}, \ch{CO}, \ch{CO2}, \ch{H2S}, \ch{HCN} for \poseidon and two additional species (\ch{CH4} and \ch{NH3}) for \pRT,  and (2) including \textit{high temperature} species that can occur at temperatures $>2000$\,K (\ch{SiO}, \ch{VO}, \ch{TiO},  \ch{AlO}, \ch{SH}); all species are listed in Table\,\ref{tab:ret_priors} along with the references to the specific line list. 

Although previous literature does not find conclusive evidence of these high-temperature species, we run an additional retrieval including them as the temperatures in the planetary atmosphere could allow for their formation. This further allows us to investigate if any features in the spectrum could be explained by these molecules, e.g., the increase in the transit depth at $\sim 5$\,\textmu m (see Fig.\,\ref{fig:transmission-spectrum_all}).  Note that we are excluding \ce{FeH} due to the absence of opacity at infrared wavelengths.

For all retrievals, we use a grey cloud deck at pressure level P$_\mathrm{cloud}$, we fit for planet gravity ($\log_{10} g$) using a Gaussian prior based on the planet's radius and mass measurements, temperature ($T$), and an offset between NRS1 and NRS2 ($\delta_\mathrm{rel}$). 
The priors of all parameters used for each of our retrieval models are given in Table \ref{tab:ret_priors} for both \poseidon and \pRT. We describe retrieval-specific setups in the following sections.

We further investigated non-isothermal P-T profiles using \poseidon in free chemistry with IR species using the \eureka R=400 data set. Specifically, we tested the Guillot profile \citep{Guillot2010OnAtmospheres} and 5-parameter profile \citep{Madhusudhan2009AAtmospheres}, but they were disfavoured by the Bayesian evidence and the structure was indistinguishable from an isothermal profile. Therefore we used an isothermal profile for all our retrievals presented here. 

In addition to our presented JWST results, we also conduct a retrieval analysis when combining NIRSpec with previously published HST data \citep{Pluriel2020ARES.WFC3,Gascon2025TheKELT-7bb} to test our conclusions. For this we use \poseidon and \nemesispy for free chemistry retrievals; we also conduct equilibrium retrievals using \nemesispy, where the abundance of \ce{H-} and \ce{e-} is a free fitting parameter, as \myplanet's HST WFC3/G141 spectrum cannot be adequately fitted with \ce{H-} at equilibrium abundances only \citep{Gascon2025TheKELT-7bb}. 
We require additional parameters in the JWST+HST retrievals to describe the optical absorption in the atmosphere. Both \poseidon and \nemesispy use a scattering slope parameterised by $\log a$ and $\gamma$ (where $a$ is the enhancement over Rayleigh scattering and $\gamma$ tunes the strength of the scattering slope), and \ce{H-} absorption. The specific retrieval setup by \poseidon and \nemesispy used for our tests using a combined JWST+HST retrievals are described in the relevant subsection for each code.

\subsubsection{\poseidon-specific setup}

To examine the atmosphere of \myplanet, we implement free chemistry retrievals with the open-source package, \poseidon \citep{MacDonald2017HDWater,MacDonald2023POSEIDON:Spectra}. The model parameter space is explored with 1000 live points, using the nested sampling package \texttt{PyMultiNest} \citep{Feroz2009MULTINEST:Physics, Buchner2014X-rayCatalogue}. 

The atmosphere is modelled as hydrogen-helium dominated with a ratio He/H = 0.17 within which trace gaseous abundances are represented by their $\log_{10}{X}$ volume mixing ratios (VMRs). We model our atmosphere at 100 pressure levels distributed uniformly in log space, with a maximum pressure of 100 bar. The reference pressure is fixed at 10 bar. 
Due to the presence of \ce{H-} in the atmosphere \citep[inferred by the analysis of][]{Pluriel2020ARES.WFC3,Gascon2025TheKELT-7bb}, we set a generous lower limit on the pressure of $10^{-10}$ bar in the free chemistry retrieval to capture the full contribution of any \ce{H-} opacity. 
The model transmission spectrum for each atmospheric model is computed at $R=30,000$ from $1 - 5.2$\,\textmu m.

For the JWST+HST retrievals, \poseidon further uses a linear wavelength grid of 10,000 points between $0.19 - 1$\,\textmu m such that the spectral resolution across this wavelength range is $> 100\times$ the resolution of the G280 data. These spectra are then convolved and binned to the resolution and wavelength solution, respectively, of our data.

\subsubsection{\texttt{petitRADTRANS}-specific setup}

We used \texttt{petitRADTRANS} (\pRT) version 3.1 \citep{Molliere2019PetitRADTRANS:Retrieval,Blain2024SpectralModel:3,Nasedkin2024AtmosphericPetitRADTRANS} to run equilibrium chemistry retrievals, which uses a pre-calculated chemistry table from \texttt{easyCHEM} \citep{Lei2024EasyCHEM:Atmospheres}, first described in \citet{Molliere2017ObservingObservations}.
Within \pRT we conduct parameter exploration using the Python version of the nested sampling algorithm \texttt{MultiNest} \citep{Feroz2009MULTINEST:Physics, Buchner2014X-rayCatalogue} where we use $500$ live points for exploring the space of the 7 total free parameters. We use 100 atmospheric layers equally distributed in log space, from 10$^{-8}$ to $10^2$ bar. The reference pressure is fixed at 0.1 bar. 

We use \pRT's correlated-$k$ opacity tables at R=1,000, for both the base species and high-temperature species. The prior ranges (in mass fractions) and references can be found in Table\,\ref{tab:ret_priors}. 
We consider \ch{H2-H2} and \ch{H2-He} collision-induced absorption \citep{Richard2012NewCIA}.

\subsubsection{\nemesispy-specific setup}
\nemesispy is a python-based version of \nemesis, a retrieval package that uses a correlated-k \citep{correlatedk} radiative transfer model combined with PyMultiNest \citep{Feroz2009MULTINEST:Physics,Feroz2008MultimodalAnalyses,Feroz2013BayesianPlanets,Buchner2014X-rayCatalogue} for the nested sampling retrieval. 
\nemesispy \citep{irwin2008,yang2024} is coupled with FASTCHEM to solve for equilibrium chemistry. 

\nemesispy has been used to analyse spectra of multiple transiting exoplanets, including KELT-7b as observed with HST \citep{Gascon2025TheKELT-7bb}. Here, we perform chemical equilibrium retrievals for HST+JWST (G280+G141, NIRSpec/G395H) with \ce{H-} and \ce{e-} abundances as additional free parameters. It is therefore perfectly equipped to use for JWST+HST combined retrievals. 

We use the 3-parameter cloud and haze model used by \cite{MacDonald2017HDWater} with an additional cloud fraction parameter. We retrieve the reference radius and pressure level of the reference radius, and offsets between instruments/detectors. For the full retrieval including the Hubble datasets, we use the 6-parameter T-p profile from \cite{Madhusudhan2009AAtmospheres}; we also include offsets for G141, NRS1 and NRS2 relative to G280, and the same stellar activity parameterization we use in Gascon et al. (under review), retrieving the photospheric temperature, the temperature of any heterogeneities, and the heterogeneity filling factor. For the NIRSpec-only retrieval, we adopt an isothermal T-p profile and retrieve the offset for NRS2 relative to NRS1, and we do not include any parameterization for stellar heterogeneities.  Note that for this investigation we used the \eureka R=400 spectrum.

\section{Results}
\label{sec:results}
We discuss our results on our NIRSpec/G395H \eureka R=400 transmission spectrum, which acts as the primary reduction in our retrieval analysis. 
To test the robustness of our free chemistry retrieval results to the choice of reduction, we also retrieve the base species and high-temperature species model on all other reductions and resolutions. While we do not discuss these separately, we note any differences in retrieval outputs when applied to different reductions or resolutions.

\subsection{Equilibrium chemistry}
\begin{figure}
    \centering
    \includegraphics[width=\columnwidth]{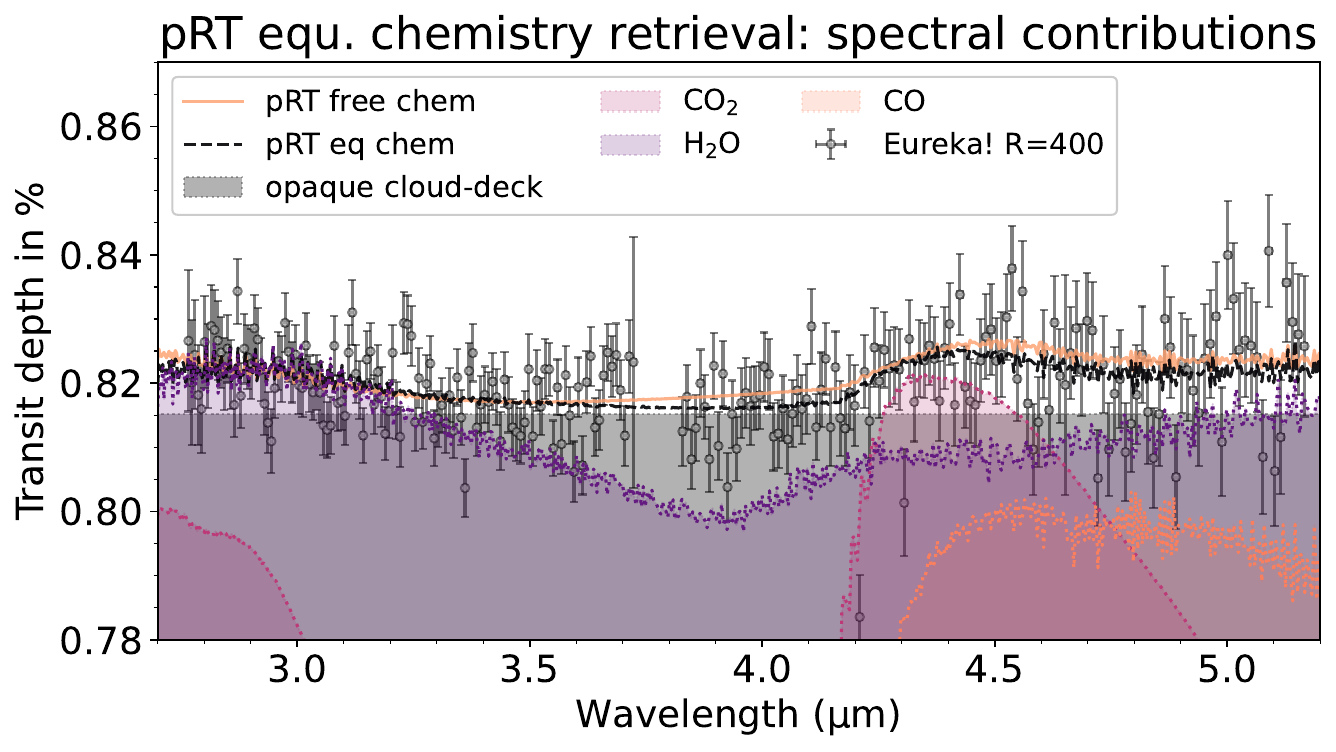}
    \caption{NIRSpec/G395H transmission spectrum of \myplanet using \eureka R=400 (black points) compared to the best-fit equilibrium (orange-dashed line) and free chemistry (magenta solid line) model using \pRT. The molecular contributions from \ce{H2O} (blue, dotted), \ce{CO2} (purple, dotted), \ce{CO} (red, dotted) to the retrieved spectrum from the best-fit equilibrium chemistry model are shown, as well as the gray cloud-deck (light gray). }
    \label{fig:prt-eq-chem-contributions}
\end{figure}

\begin{figure}
    \centering
    \includegraphics[width=\linewidth]{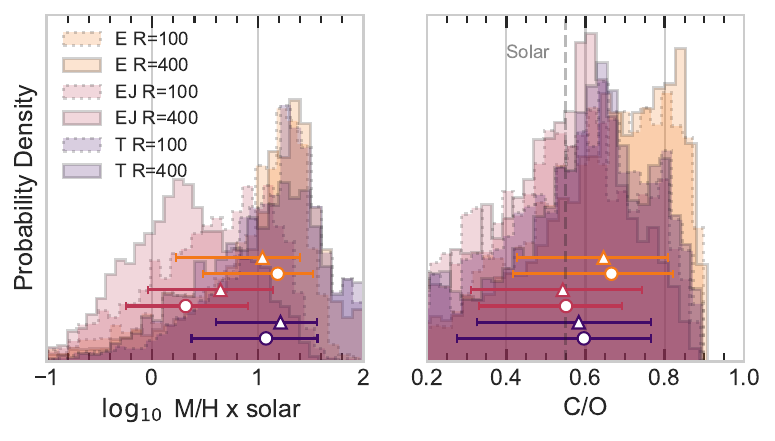}
    \includegraphics[width=\linewidth]{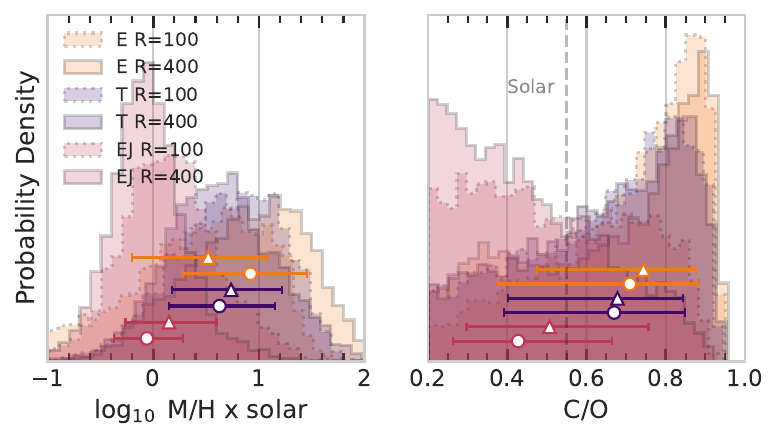}
        \caption{Retrieved C/O and metallicity for \myplanet from
the \pRT equilibrium chemistry retrievals (top) and \poseidon (bottom), showing each of the reductions (\eureka (E), \exoticjedi (EJ), \tiberius (T)) at
both R=100 and R=400 spectral resolutions. The markers (R=100: triangle; R=400: round) correspond to the median value, the errors to the 16th and 84th percentile.}
    \label{fig:ctoo-metallicities-eq}
\end{figure}

\myplanet's atmosphere retrieved using \pRT and \poseidon chemical equilibrium setups shows wide ranges for the atmospheric C/O and metallicity. The best-fit and the opacity contributions are shown in Fig.\,\ref{fig:prt-eq-chem-contributions}. The C/O ratios retrieved for all reductions and resolutions demonstrate a range from 0.43--0.74, with relatively large uncertainties of up to 0.32 and do not exhibit a Gaussian shape, see Fig.\,\ref{fig:ctoo-metallicities-eq} and Table\,\ref{tab:CtoO_mets_eq_chem}. They all agree well within the $1\sigma$ for that reason and no direct constraints can be placed on the atmospheric C/O ratio of \myplanet. Though, we note that all posteriors from the reductions consistently drop off sharply at a C/O ratio greater than $0.9-1$ at the $\sim2\sigma$ level.

\begin{figure}
    \centering
    \includegraphics[width=0.9\linewidth]{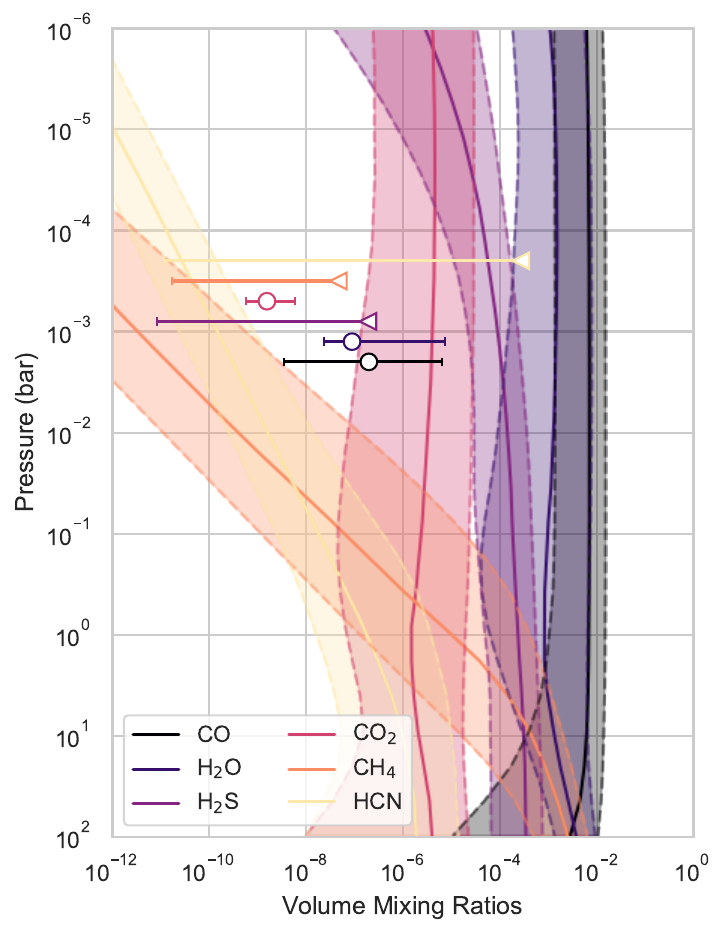}
    \caption{Pressure and volume mixing ratios; a comparison between the retrieved values through free chemistry (points with $1\sigma$ uncertainties for CO, \ce{H2O} and \ce{CO2} and $2\sigma$ range as an upper limit for the remaining unconstrained species) and the profiles calculated based on the retrieved C/O ratio and metallicity from the equilibrium chemistry model (using their $1\sigma$ uncertainties), both with \pRT. The individual colours correspond to the molecules included in the model. }
    \label{fig:vmrs-pressures-eq-vs-free-chem}
\end{figure}

The metallicities retrieved in the equilibrium chemistry run for both \pRT and \poseidon also display large uncertainties and their median values range from approximately solar to $\sim 15\times$ solar. The retrieval codes vary slightly in their metallicities, \pRT preferring higher metallicities than \poseidon, however, similar to the C/O ratio this is all within the $1-2\sigma$ uncertainties as the spectrum does not allow for tighter constraints. In Fig.\,\ref{fig:vmrs-pressures-eq-vs-free-chem}, we show the computed corresponding range of VMRs for each molecule based on the retrieved C/O ratios and metallicities for \myplanet's atmosphere.

We find the best-fit to be an atmosphere with a high-altitude cloud-deck, consistent across reductions, with a tail towards lower altitudes. The wide, unconstrained nature of the cloud-deck for both \poseidon and \pRT likely contributes to the large uncertainties in C/O and metallicity as any abundances are challenging to constrain if the amount of cloud opacity is unknown. This is demonstrated by the fact, that this high-altitude cloud layer by the equilibrium chemistry retrievals is contrary to our results from free chemistry models with low-altitude clouds (see Section\,\ref{sec:results-free-chemistry}).  The temperature also shows a large range of possible solutions with the $1\sigma$ contours spanning a range of $\sim 1,000$\,K. Cloud deck layer and temperature show a slight correlation in our posteriors, where the higher cloud deck altitudes are associated with higher temperatures ($>1300$\,K), more consistent with \myplanet's equilibrium temperature. The posterior plots for both retrievals and all three reductions at R=400 are displayed in the Appendix, Fig.\,\ref{fig:corner-eq-chem-r400-nirspec-only-pos} \& Fig.\,\ref{fig:corner-eq-chem-r400-nirspec-only} for \poseidon and \pRT, respectively.

\begin{table}
\def\arraystretch{1.5}
\caption{Retrieved C/O ratios and metallicities of \myplanet's atmosphere in chemical equilibrium. Solar refers to \citet{Asplund2009TheSun}. }
    \label{tab:CtoO_mets_eq_chem}
    \centering
    \begin{tabular}{lcc}
    \hline
     \multicolumn{3}{c}{\textbf{Equilibrium chemistry:} \pRT} \\  \hline
     Reduction & C/O  & Z ($\times$solar) \\ \hline
     \eureka R=100    & $0.65^{+0.17}_{-0.22}$  & ${11.1}^{+14}_{-9.4}$ \\
     \eureka R=400   & $0.67_{-0.25}^{+0.16}$ & ${15}^{+18}_{-13}$  \\
     \tiberius R=100    & $0.58_{-0.26}^{+0.19}$ & ${16}^{+20}_{-13}$ \\
    \tiberius R=400     & $0.59_{-0.32}^{+0.17}$ & ${11.9}^{+25}_{-9.6}$  \\
     \exoticjedi R=100    & $0.54_{-0.24}^{+0.20}$ & ${4.4}^{+9.6}_{-3.6}$ \\
     \exoticjedi R=400    & $ 0.55_{-0.22}^{+0.15}$ & ${2.1}^{+6.0}_{-1.6}$ \\ 
     \hline
    \hline
     \multicolumn{3}{c}{\textbf{Equilibrium chemistry:} \poseidon} \\  \hline
     Reduction & C/O  & Z ($\times$solar) \\ \hline
     \eureka R=100    & $0.74^{+0.13}_{-0.27}$ & ${3.2}^{+17.0}_{-0.9}$ \\
     \eureka R=400   &  $0.71^{+0.17}_{-0.33}$ & ${8.3}^{+37.2}_{-2.5}$ \\
     \tiberius R=100    & $0.68^{+0.17}_{-0.28}$ & ${5.5}^{+20.0}_{-1.8}$ \\
    \tiberius R=400     & $0.67^{+0.18}_{-0.28}$ & ${4.2}^{+12.6}_{-1.2}$\\
     \exoticjedi R=100    & $0.51 ^{+0.25}_{-0.21}$ &  ${1.4}^{+3.5}_{-0.5}$\\
     \exoticjedi R=400    & $0.43^{+0.24}_{-0.16}$ & ${0.9}^{+1.8}_{-0.4}$\\ 
     \hline
    \end{tabular}
\end{table}      

\subsection{Free chemistry}
\label{sec:results-free-chemistry}

\subsubsection{Base model}
The free chemistry base model (only including near-infrared species, see Section\,\ref{sec:retrievals}) retrieves low abundances for \ch{H2O} and \ch{CO2} using both \poseidon and \pRT. Best-fit temperatures for \poseidon and \pRT vary: while the former retrieves a temperature of $\sim1600$\,K, \pRT favours a temperature $\sim 2,000$\,K, though both are poorly constrained with uncertainties of $>500$\,K. In \poseidon, we also find a bimodal temperature distribution, where a very low temperature ($<900$\,K, pushing the lower prior bound), high mean molecular weight solution, with high abundance modes of \ce{CO} and \ce{CO2} (pushing the upper prior boundaries) also able to provide a good fit to the data. These high atmospheric \ce{CO} and \ce{CO2} fractions of the order of 0.1 cannot be explained physically for a hot Jupiter of this density, nor does the temperature mode correspond to a realistic limb average temperature for a $T_\mathrm{eq}$ = 2048\,K planet. This mode is not correlated with the remaining species in the model. Volume mixing ratios (VMRs) of \ce{H2O}, \ce{CO2} and \ce{CO} and detection significances using Bayesian evidence differences for both retrievals are listed in Table\,\ref{tab:species_det_sig}. 

\begin{figure}
    \centering
    \includegraphics[width=\linewidth]{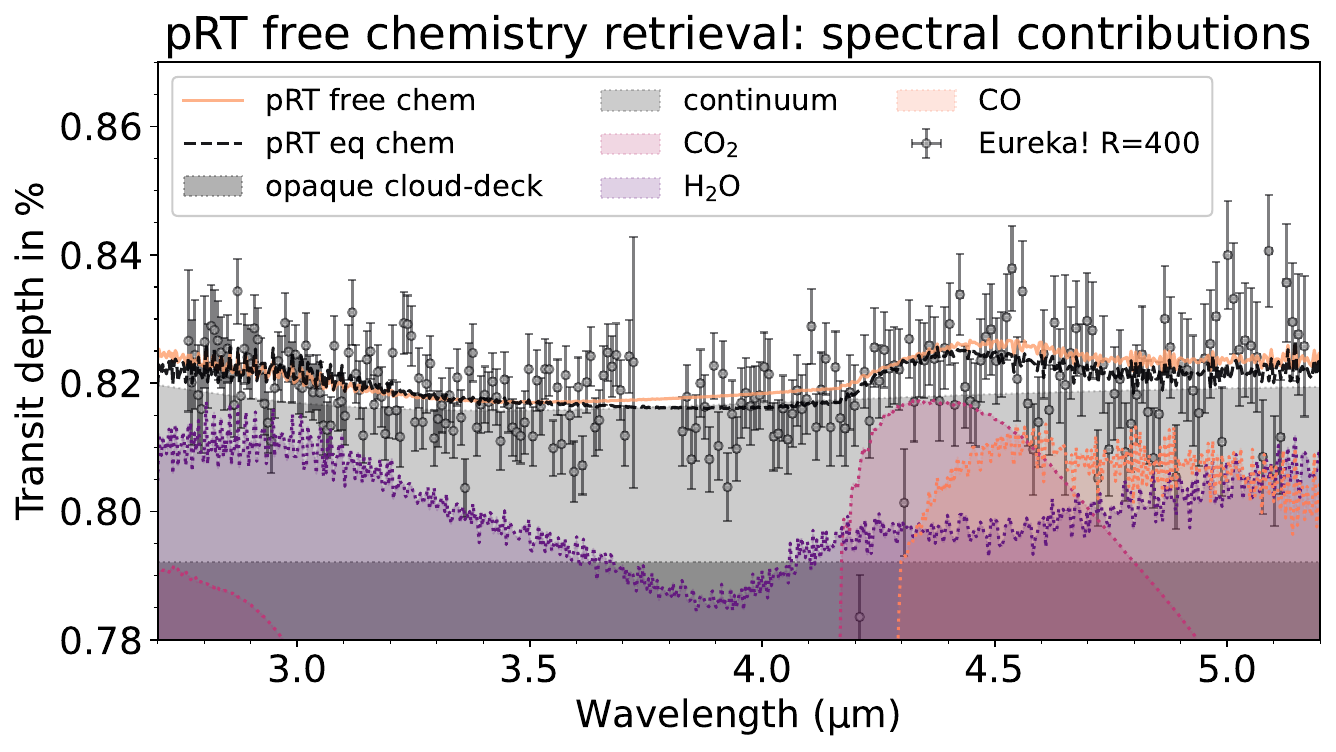}
    \caption{Similar to Fig.\,\ref{fig:prt-eq-chem-contributions}, NIRSpec/G395H transmission spectrum of \myplanet using \eureka R=400 (black points) compared to the best-fit equilibrium (orange-dashed line) and free chemistry (magenta solid line) model using \pRT. Here the molecular contributions (median) are shown from the free chemistry retrieval: \ce{H2O} (blue, dotted), \ce{CO2} (purple, dotted), \ce{CO} (red, dotted) and the gray cloud-deck (light gray). The free chemistry retrievals (both \poseidon and \pRT) find a low altitude cloud deck and low abundance of infrared species, probing the continuum opacity level, thus, we also include the continuum's contribution (off the plot axis in the equilibrium chemistry case).} 
    \label{fig:prt-free-chem-contributions}
\end{figure}

\begin{table}
\def\arraystretch{1.2}
\caption{Retrieved volume mixing ratios (VMRs) for \ce{H2O}, \ce{CO2} and \ce{CO} from our \poseidon and \pRT retrievals, where the uncertainties refer to the $1\sigma$ confidence intervals. The detection significances for each detected species are shown here as the differences in Bayesian evidence values $\Delta \ln \mathcal{Z}$, where $1.0 < \Delta \ln \mathcal{Z} < 2.5$ is considered weak to tentative evidence for a detection and $\Delta \ln \mathcal{Z} < 1.0$ is considered inconclusive \citep{Jeffreys1983TheoryProbability}. The model including CO for \poseidon is not preferred at all as $\Delta \ln \mathcal{Z} < 0$.}
    \label{tab:species_det_sig}
\begin{tabular}{lccc}
    \hline 
     \eureka R=400 & \ce{H2O} & \ce{CO2} & \ce{CO} \\ \hline
    \poseidon \\ 
    log$_{10}$(VMR) & $-7.05^{+0.67}_{-1.96}$ & $-8.46^{+6.02}_{-0.28}$ & $-7.23^{+4.85}_{-2.76}$\\ 
    $\Delta \ln \mathcal{Z}$ & $1.9$ & $1.5$ & $-0.7$  \\
    \hline
    \pRT \\  
    log$_{10}$(VMR)  & $-7.05^{+0.59}_{-1.93}$ & $-8.81^{+0.44}_{-0.58}$  & $-6.7^{+1.8}_{-1.6}$\\ 
    $\Delta \ln \mathcal{Z}$ & $1.0$ & $ 0.9$ & $0.3$ \\
\hline
    \end{tabular}   
\end{table}

\poseidon retrieves an atmosphere that is cloud-free, with a lower cloud deck pressure level at $\log_{10} \mathrm{P_{cloud}} = 0.63^{+0.90}_{-0.94}$\,bar. \pRT also retrieves this solution for the cloud-deck (at $\log_{10} \mathrm{P_{cloud}} \sim 0.38$), but an additional mode is also seen at  $\log_{10} \mathrm{P_{cloud}} \sim -3.5$ which is consistent with the preferred cloud-top pressure by the equilibrium chemistry models.  The combination of a low altitude cloud deck and low abundance of infrared species means that the spectrum is probing the continuum opacity level, clearly visible in Fig.\,\ref{fig:prt-free-chem-contributions} which displays the molecular as well as continuum contributions to the transmission spectrum. 

Each reduction finds broadly consistent parameters to the \eureka R=400 reduction --- all atmospheric parameters fall within 1 sigma of each other. The high mean molecular weight, low temperature mode is only visible using \poseidon and is most prominent for the \eureka R=400 reduction, though still present to a lesser extent in the remaining reductions. The high altitude cloud-top pressure mode is mostly also recovered in the other reductions when using \pRT; only \exoticjedi R=400 does not show a peak in the posterior distribution in that parameter space. The posterior distributions of this base model using free chemistry for both \poseidon and \pRT with the different reduction can be found in the Appendix, Fig.\,\ref{fig:corner-free-chem-r400-nirspec-only-pos} and Fig.\,\ref{fig:corner-free-chem-r400-nirspec-only}, respectively.

\begin{figure*}
    \centering
    \includegraphics[width=\linewidth]{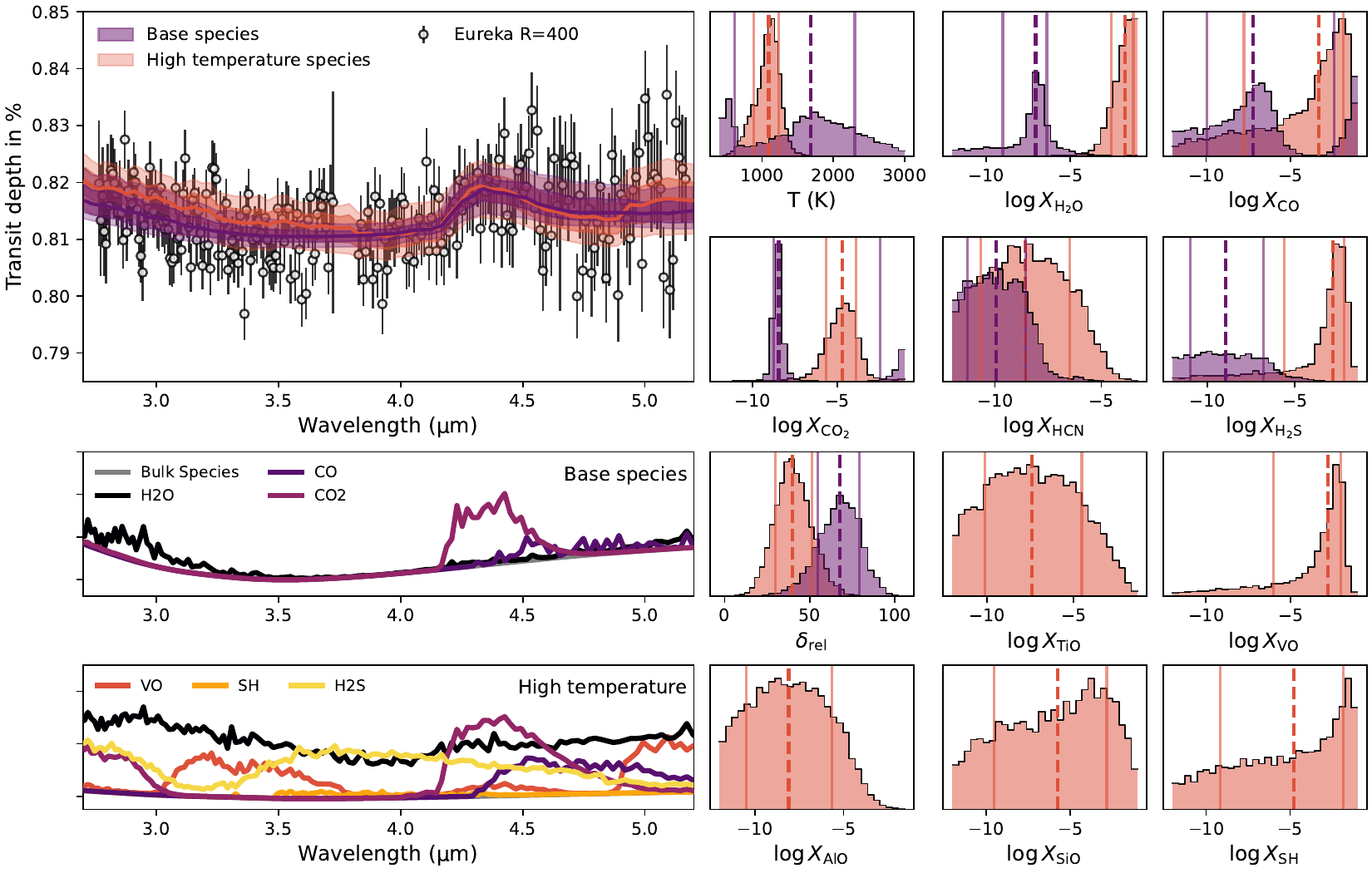}
    \caption{Comparison between the free chemistry retrievals using the \textit{base} species as well as including \textit{high-temperature} species. Top left panel: JWST NIRSpec/G395H transmission spectrum of KELT-7b using \eureka R=400 and the two competing free chemistry retrievals (base species and high-temperature species). Bottom left panels: Spectral contributions of the fitted species for the two models considered. Right panels: posterior distributions of the fitted retrieval parameters.  
    While the high temperature species are statistically favoured it produces unrealistically high abundances for a number of species and thus we rule it out for KELT-7b.
    }
    \label{fig:nirspec-base-vs-high-T}
\end{figure*}

\subsubsection{Model with high-temperature species}
The spectrum retrieved by \poseidon when including the high-temperature species and the retrieved parameters are displayed together with the retrieval with only the base species in Fig.\,\ref{fig:nirspec-base-vs-high-T}. For both \poseidon and \pRT, the atmosphere that is retrieved produces a slightly smaller offset ($15-20$\,ppm smaller) between NRS1 and NRS2. It is characterised by high abundances of all molecules, specifically \ce{H2S}, \ce{CO} and \ce{VO}, with \ce{H2O} and \ce{SH} pushing toward the upper prior bound. The spectral decomposition (bottom left, Fig.\,\ref{fig:nirspec-base-vs-high-T}) shows that the flattening of the \ce{H2O} slope is being produced by the high \ce{H2S} abundance, exhibiting the same shape as a low \ce{H2O} abundance and continuum opacity. At the edge of the NRS2 detector, around 5\,\textmu m, the transit depths increase which is fit by opacity from the high \ce{VO} ($\log_{10}$ VMR $= -2.79^{+0.77}_{-3.23}$), which is physically unrealistic. 

\pRT shows a statistical preference for the base species model as the Bayesian evidence for the model including the high-temperature species decreases by $\Delta \ln \mathcal{Z} = 2.0$. In contrast, \poseidon statistically prefers the high temperature species model with $\Delta \ln \mathcal{Z} = 2.44$. However, it is unlikely that such species could exist at the high abundances inferred by the retrieval and they are not detected at optical wavelengths in \citet{Gascon2025TheKELT-7bb} where their opacity should be most prominent. Therefore, with \pRT finding no evidence for high-T species and \poseidon producing a physically implausible atmosphere, we conclude it is unlikely that any molecules that form at the high temperature regime are contributing to the opacity in \myplanet's NIRSpec transmission spectrum. 

\subsubsection{Combined with HST}
To be able to distinguish between the two preferred models, we investigate whether the previously published two HST data sets in the UV and IR can aid in constraining the continuum and/or cloud layer, as well as the \ce{H2O} abundance. However, as the wavelength ranges of none of the data sets overlap, the offset between the spectra is unknown. In addition, the \ce{H-} in the atmosphere of \myplanet and the lack of knowledge of abundance constraints on said \ce{H-} prohibit any inference about the continuum or cloud level. 

\begin{figure*}
    \centering
    \includegraphics[width=\textwidth]{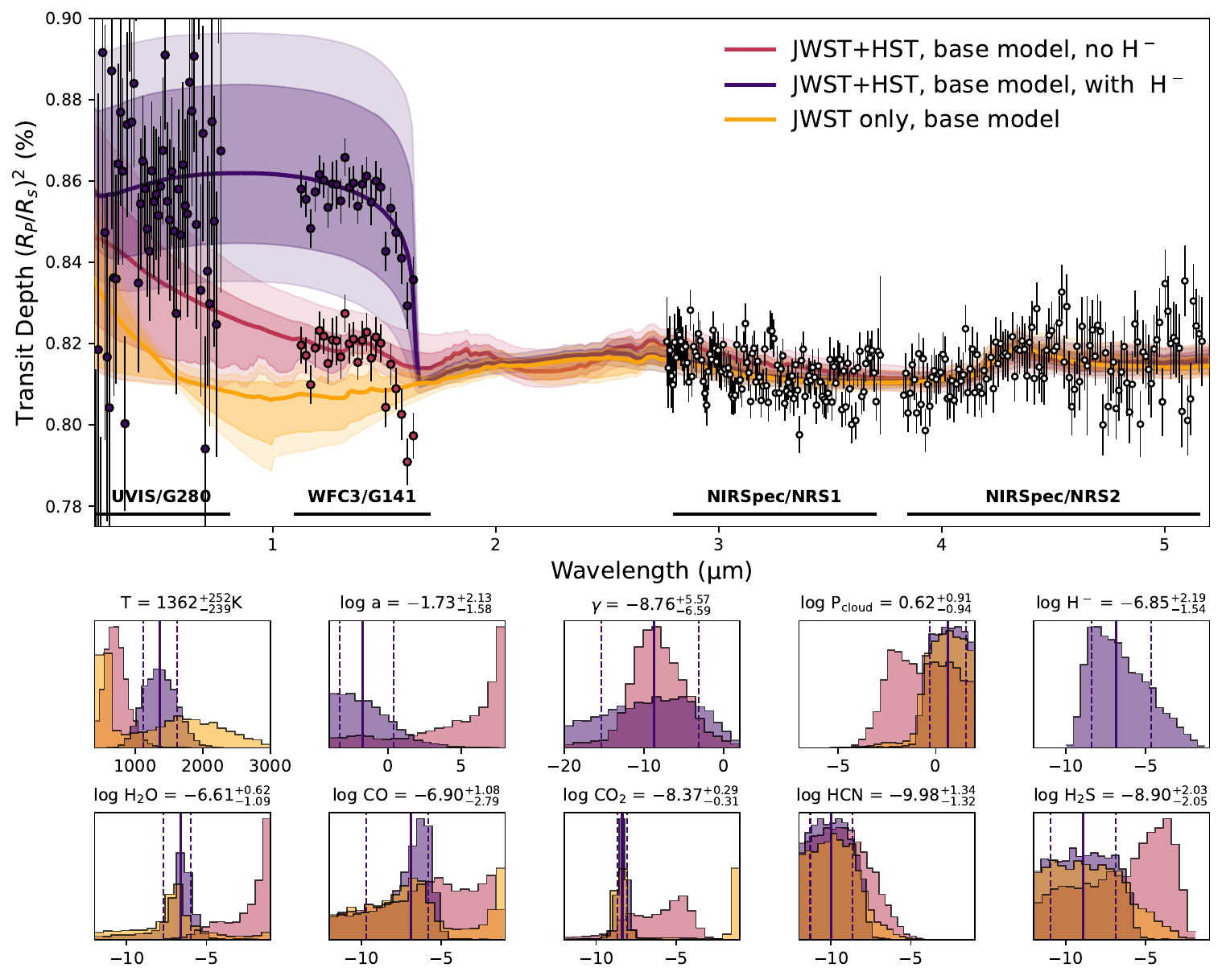}
    \caption{Top: \poseidon free chemistry retrievals comparison between using only JWST data (orange) and in combination with previously published HST data (pink: excluding \ce{H-}; purple: including \ce{H-}). For clarity, we include the WFC/G141 data twice, as the difference in offset retrieved when including \ce{H-} is significant. The colour of the data corresponds to the offset applied from each model. For the UVIS/G280 data we only applied one offset as the uncertainties are relatively large and the offset difference is not as significant. \newline Bottom: The posterior plots for the retrieved species, cloud deck and temperature ($T$), as well as scattering slope parameters for the retrievals with HST data ($\log a, \gamma$). The colours correspond to the individual models, demonstrating that the inclusion of HST data does not provide further constraints for the cloud deck or abundances as the unknown \ce{H-} abundance in combination with the lack of overlapping wavelengths prevents inferences about the continuum of \myplanet's atmosphere. Quoted median values and 1$\sigma$ errors correspond to the posteriors retrieved on the JWST and HST data when including \ce{H-} opacity, shown with the vertical solid and dashed lines.}
    \label{fig:appendix_g395h_hst}
\end{figure*}

Therefore, our free chemistry retrievals with JWST+HST combined is not able to provide further constraints on the atmospheric composition of \myplanet. We find good agreement in the retrieved atmospheric parameters in the free chemistry model, see Fig.\,\ref{fig:appendix_g395h_hst}. Similarly, the metallicity and C/O ratio in the equilibrium chemistry model with free \ce{H-} and \ce{e-} is  consistent and shows equally wide posteriors compared to the NIRSpec-only retrieval, see Appendix, Fig.~\ref{fig:appendix_g395h_hst_nemesispy}.

\section{Discussion \& Conclusions}
\label{sec:discussion-conclusions}

We present the transmission spectrum of the aligned hot Jupiter \myplanet obtained from a single transit observation with JWST NIRSpec/G395H, covering the 2.8--5.2\,\textmu m wavelength ranges. We find evidence for a mirror-tilt event in our data, also visible in the JWST fine guidance data. The time-series spectra were reduced using three independent pipelines, which showed offsets between each other, likely a combination of the uncertainty in flux change due to the mirror-tilt event and the low number of groups/integration (5) causing higher noise. 

We find that our transmission of \myplanet shows only weak features of \ce{H2O} and \ce{CO2}.
To draw inferences about \myplanet's atmosphere based on our NIRSpec/G395H spectrum, we use two independent atmospheric retrieval codes, \pRT and \poseidon, and apply equilibrium and free chemistry setups. With our retrieval setups, we find two competing scenarios for the atmosphere of \myplanet: (1) a cloud layer at high altitude and small features, or (2) a cloud-free low-metallicity atmosphere where we probe the continuum layer. Since the simpler, equilibrium chemistry model is statistically preferred (see $\ln \mathcal{Z}$ in Table\,\ref{tab:full-retrieval-results}) in all reductions, resolutions and retrieval setups at $\Delta$ $\ln \mathcal{Z} = 1.6--3.1$, the slightly preferred scenario is that \myplanet's atmosphere has a (1) relatively high cloud-deck which masks the molecular features. Based on \myplanet's equilibrium temperature of $2050$\,K, cloud condensate species such as \ce{SiO2} or \ce{Al2O3} can form and explain the high-altitude cloud deck \citep[e.g., see][]{Wakeford2015TransmissionExoplanets, Parmentier2016TRANSITIONSJUPITERS}.  We further tested cloud fraction as a free parameter to allow for patchy clouds, though it was not statistically preferred (Bayesian evidence difference at $\Delta \ln(\mathcal{Z}) \sim 1$). We note the inclusion of these patchy clouds resulted in a preference for a cloud deck at higher altitude and a higher metallicity atmosphere.  

We investigate evidence for \ce{H2O}, \ce{CO2} and \ce{CO} in our transmission spectrum of \myplanet. However, we are not able to detect strong evidence for any of the molecules using our free chemistry setup. We attribute this to the small feature sizes, the unknown cloud layer and the unknown NRS1/NRS2 offset that we need to fit for. 
We also find a potential feature in our transmission spectrum at $\sim 5$\,\textmu m that could be of atmospheric origin. We investigated species that may form at the higher temperatures possible on the dayside of \myplanet and our free chemistry retrieval suggests \ce{VO} as a possible absorber. However, the retrieved abundance is relatively high. In addition, \ce{VO} has not been detected in \myplanet's optical transmission spectrum \citep{Gascon2025TheKELT-7bb} and therefore unexpected to be detected in \myplanet's NIRSpec/G395H spectrum. 

We extend our free chemistry retrieval analysis to HST wavelength ranges, as HST UVIS/G280 \citep{Gascon2025TheKELT-7bb} and WFC3/G141 \citep{Pluriel2020ARES.WFC3} transmission spectra are available for \myplanet. However, even the addition of these data did not result in a clear indication of the cloud layer due to the unknown H- opacity and the fact that we are unable to constrain the offset between the transmission spectra from the different detectors (see Fig.\,\ref{fig:appendix_g395h_hst}). We find similar conclusions when running equilibrium chemistry retrievals with free H-/e-; the posteriors between the JWST only and JWST+HST combined retrieval are nearly identical (see Fig.\,\ref{fig:appendix_g395h_hst_nemesispy}). 

Due to the lack of strong features, we are unable to place meaningful constraints on the C/O ratio of \myplanet's atmosphere, though all reductions show sharp drop at C/O ratios $>0.9-1$ at a $\sim 2\sigma$ level. Due to the evidence for \ce{H-} in the HST wavelength range, it is likely that any modelling to measure C/O needs to take the amount of water dissociation in the upper atmosphere into account. Therefore, we are unable to quote a C/O ratio for the atmosphere of \myplanet. As a result, this does not place strong constraints on \myplanet's history as the C/O ratio we retrieve of $0.25$--$0.9$ is consistent with many possible formation scenarios.

The retrieved metallicity of \myplanet's atmosphere compared to solar, $[Z/H]=-0.2$--$1.5$ dex, is likely stellar or super-stellar: the star \mystar has a metallicity of [Fe/H]=$0.139^{+0.075}_{-0.081}$ dex \citep{Bieryla2015}. A super-stellar planetary metallicity can be created by various processes. These include an enhancement of any solid accretion through pebbles or planetesimals \citep[e.g.][]{Danti2023,Penzlin2024BOWIE-ALIGN:Compositions}; or the planet accreting significant fraction of its gaseous envelope at a location and time where the gas-phase metallicity is enhanced, e.g., near an ice line where volatiles are preferentially released from drifting pebbles \citep{Booth2017ChemicalDrift,Bitsch2023} or in a region where the metallicity is enhanced by disc photoevaporation \citep{Lienert2024}. 
While this single planet does not constrain any specific scenario, it is an important piece in the puzzle to construct a comprehensive sample of similar planets in the BOWIE-ALIGN programme. The aim the programme, to probe whether the atmospheric compositions of aligned (disc-migrated) and misaligned (high-e migrated) hot Jupiters differ significantly, may only succeed when studying a population instead of individual planets \citep{Kirk2024BOWIE-ALIGN:History}.

We caution that one-transit observations with NIRSpec/G395H may not be sufficient to constrain \ce{H2O} and \ce{CO2} in the atmosphere of a hot Jupiter which may exhibit temperatures closer to ultra-hot Jupiters where \ce{H2O} dissociation takes an effect as demonstrated by this study. Ultra-hot Jupiters observed with JWST include WASP-121b (T$_\mathrm{eq} \sim  2300$\,K), where observations with NIRSpec/G395H also find weak evidence for \ce{H2O} and no evidence for \ce{CO2} in the atmosphere, though strong SiO and CO features \citep{Gapp2025WASP-121Atmosphere}. \citet{Lothringer2025RefractoryJWST} find a NIRSpec/G395H spectrum similar to WASP-121b for ultra-hot Jupiter WASP-178b (T$_\mathrm{eq} \sim  2500$\,K; weak \ce{H2O}, no \ce{CO2}, detection of CO and SiO). In contrast to \myplanet, \citet{Lothringer2025RefractoryJWST} find that meaningful abundance constraints were possible when the JWST data was combined with the short-wavelength HST observations as WASP-178b shows strong absorption in the UV ($<0.3$\textmu m).

This provides a challenge when trying to constrain C/O ratios and atmospheric metallicity of ultra-hot Jupiters solely from the near-infrared wavelength ranges. Wider wavelength observations hold the key to constraining the cloud layer and determining the chemical composition present in \myplanet's atmosphere. However, obtaining overlapping wavelength ranges is crucial. As is the case with our study of \myplanet, adding HST UVIS/G280 and WFC3/G141 data to NIRSpec/G395H may not be sufficient to disentangle the continuum from other atmospheric processes (in this case \ce{H-}). Both additional optical and mid-infrared wavelength ranges will be valuable in constraining the cloud-deck layer, the \ce{H-} contribution, and ultimately the abundance of carbon- and oxygen-bearing molecules.

\section*{Acknowledgements}
We thank the anonymous referee for their time to review this manuscript and for their valuable comments that improved this manuscript.

This work is based on observations made with the NASA/ESA/CSA JWST. The data were obtained from the Mikulski Archive for Space Telescopes at the Space Telescope Science Institute, which is operated by the Association of Universities for Research in Astronomy, Inc., under NASA contract NAS 5-03127 for JWST. These observations are associated with program \#3838. This work was inspired by collaboration through the UK-led BOWIE+ collaboration. Support for program JWST-GO-3838 was provided by NASA
through a grant from the Space Telescope Science Institute, which is operated by the Association of Universities for Research in Astronomy, Inc., under NASA contract NAS 5-03127. 

We are grateful to the HUSTLE programme team  (GO-17183, PI: Wakeford) for providing us the HST UVIS/G280 transmission spectrum of KELT-7b before publication of their manuscript.

JK acknowledges financial support from Imperial College London through an Imperial College Research Fellowship grant. HRW was funded by UK Research and Innovation (UKRI) framework under the UK government’s Horizon Europe funding guarantee for an ERC Starter Grant [grant number EP/Y006313/1].
RAB and AP thank the Royal Society for their support in the form of a University Research Fellowship. AP acknowledges funding from the European Union under the European Union's Horizon Europe Research and Innovation Programme 101124282 (EARLYBIRD). Views and opinions expressed are, however, those of the authors only and do not necessarily reflect those of the European Union or the European Research Council. Neither the European Union nor the granting authority can be held responsible for them.
PJW acknowledges support from the UK Science and Technology Facilities Council (STFC) through consolidated grant ST/X001121/1.
NJM, DES, and MZ acknowledge support from a UK Research and Innovation (UKRI) Future Leaders Fellowship (Grant MR/T040866/1), a Science and Technology Facilities Funding Council Nucleus Award (Grant ST/T000082/1), and the Leverhulme Trust through a research project grant (RPG-2020-82).
JEO is supported by a Royal Society University Research Fellowship. This project has received funding from the European Research Council (ERC) under the European Union’s Horizon 2020 research and innovation programme (Grant agreement No. 853022). JKB is supported by a UK Science and Technology Facilities Coundil Ernest Rutherford Fellowship (ST/T004479/1).

\section*{Data Availability}
The data are available on the Mikulski Archive for Space Telescopes at the Space Telescope Science Institute, under program number \#3838. The specific observations analysed can be accessed via \url{https://doi.org/10.17909/a3r9-tv81}. The data products and posterior plots associated with this manuscript are available on Zenodo via \url{https://doi.org/10.5281/zenodo.17092001}.
 



\input{references.bbl}
\bibliographystyle{mnras}




\appendix

\section{Transmission spectrum from combined NRS1 and NRS2 light curve fit}
\label{sec:appendix_WL_combined}
\begin{figure}
    \includegraphics[width=\columnwidth]{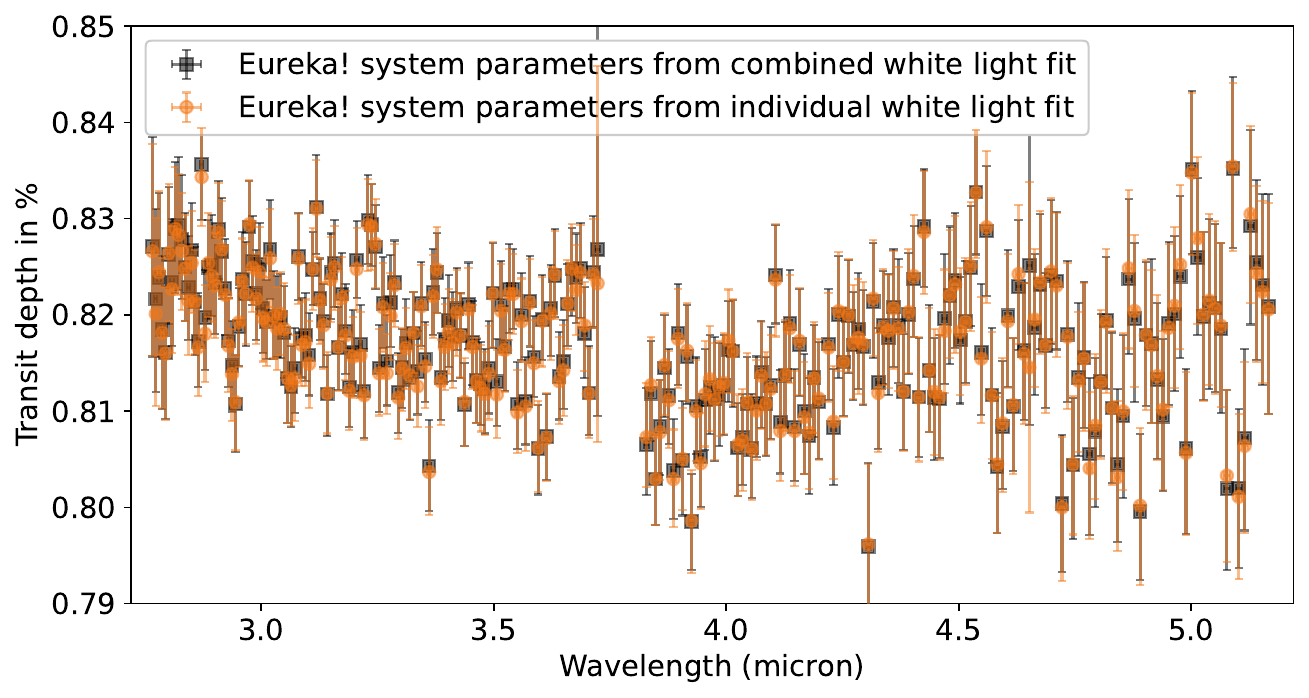}
    \caption{Comparison of two JWST NIRSpec/G395H transmission spectrum of \myplanet using \eureka R=400 light curves, one using the system parameters fitted using a combined fit of the two NRS1 and NRS2 white light curves and one using the system parameters individually retrieved from the NRS1 and NRS2 white light curves. The differences between the two are minor, the transit depths when using the combined fit system parameters are on average $<1$ppm and $<5$ppm lower for NRS1 and NRS2, respectively, compared to the transit depths using individually fit system parameters. For comparison, the average uncertainties on the transit depth in NRS1 (50ppm) and NRS2 (69ppm) are around a factor of seven higher and the fitted offset in the retrieval for \eureka R=400 is of order 50--70ppm.   }
    \label{fig:appendix-eureka-spectrum-comparison-combined-fit}
\end{figure}

To investigate whether any offsets in the transit depth between NRS1 and NRS2 could be caused by differences in system parameters, we conducted a combined white light curve fit between NRS1 and NRS2. We further refitted the \eureka R=400 light curves using the system parameters obtained by the combined fit and compared to the transmission spectrum obtained when using individual system parameters. The resulting transmission spectra are shown in Fig.\,\ref{fig:appendix-eureka-spectrum-comparison-combined-fit}. The effects are minor, on average $<5$\,ppm, compared to the offset found by the retrievals of up to 70\,ppm.

\section{Retrieval prior ranges}
\label{sec:appendix_retrievals_priors}
In Tables\,\ref{tab:ret_priors}\&\ref{tab:nemesispy_priors}, we show the prior ranges of the retrieval parameters for \pRT, \poseidon and \nemesispy (for combined HST+JWST).

\begin{table}
    \centering
    \caption{Parameters and prior ranges for our retrieval of \myplanet \poseidon, \pRT and \nemesispy (in continued Table\,\ref{tab:nemesispy_priors}). Note that they are all uniform unless stated otherwise. The rightmost column refers to the literature references of the line lists as follows: [1] \citet{Polyansky2018ExoMolWater},[2] \citet{Yurchenko2020ExoMolCO2}, [3] \citet{li2015h2o}, [4] \citet{yurchenko2024ch4}, [5] \citet{Coles2019ExoMolAmmonia},[6] \citet{Barber2014ExoMolHNC}, [7] \citet{Azzam2016ExoMolH2S} [8] \citet{yurchenko2024ch4}, [9]\citet{mckemish2019tio}, [10] \citet{mckemish2016vo}, [11] \citet{patrascu2015alo},[12] \citet{gorman2019sh}, [13] \citet{john1988}}
\label{tab:ret_priors}
    \begin{tabular}{lccc}
    \hline
    \multicolumn{4}{c}{\texttt{POSEIDON}} \\ \hline
      & \textbf{Parameter}             & \textbf{Prior Range} &  \textbf{Reference}
      \\ \hline

\textbf{Equilibrium}   & $\mathrm{C/O}$  & 0.2 -- 1.2 \\
\textbf{chemistry}  & $\log(\mathrm{M/H})$ & -1 -- 2 \\
\hline
\textbf{Free chemistry}  & $\log(\mathrm{H_2O})$ & -12 -- -1 & [1] \\
\textit{base species}            & $\log(\mathrm{CO_2})$ & -12 -- -1
 & [2]\\
(VMR)            & $\log(\mathrm{CO})$ & -12 -- -1 & [3]\\
& $\log(\mathrm{CH_4})$ & -12 -- -1 & [4] \\
            & $\log(\mathrm{NH_3})$ & -12 -- -1 & [5]\\
            & $\log(\mathrm{HCN})$ & -12 -- -1 & [6] \\
            & $\log(\mathrm{H_2S})$ & -12 -- -1 & [7]\\ \hline
\textbf{Free chemistry}            & $\log(\mathrm{SiO})$ & -12 -- -1 & [8] \\
\textit{high-temperature}           & $\log(\mathrm{TiO})$ & -12 -- -1 & [9]\\
\textit{species}            & $\log(\mathrm{VO})$ & -12 -- -1 & [10] \\
(VMR)              & $\log(\mathrm{AlO})$ & -12 -- -1 & [11] \\
            & $\log(\mathrm{SH})$ & -12 -- -1 & [12]\\
\hline

\hline

Temp. Profile & $T$ (K) & 400 -- 3000 \\
        
\hline

Ref. Radius & $R_\mathrm{ref} ( \mathrm{R_p})$ &  $0.85 - 1.15$ \\
Planet gravity  & $\log{g}$  & \multicolumn{2}{l}{$\mathcal{N}(\log{g_\mathrm{p}}, \log{g_\mathrm{p, err}})$} \\

\hline
Clouds  & $\log P_\mathrm{cloud} (\mathrm{bar})$ & -7 -- 2\\

\hline
Offset &  $\delta_\mathrm{rel}$ (ppm) & -1000 -- 1000 \\ \hline

    \hline
    \multicolumn{4}{c}{\texttt{petitRADTRANS}} \\ \hline
      & \textbf{Parameter}             & \textbf{Prior Range} & \textbf{Reference}\\ \hline

\textbf{Equilibrium}   & $\mathrm{C/O}$  & 0.1 $-$  1.5 \\
\textbf{chemistry}  & $\log(\mathrm{M/H})$ & -1 $-$  2 \\
\hline
\textbf{Free chemistry}  & $\log(\mathrm{H_2O})$ & -10 -- -1e-6 & [1] \\
\textit{base species}            & $\log(\mathrm{CO_2})$ & -10 -- -1e-6
 & [2]\\
(mass fractions)            & $\log(\mathrm{CO})$ & -10 -- -1e-6 & [3]\\
& $\log(\mathrm{CH_4})$ & -10 -- -1e-6 & [4] \\
            & $\log(\mathrm{NH_3})$ & -10 -- -1e-6 & [5]\\
            & $\log(\mathrm{HCN})$ & -10 -- -1e-6 &  [6]\\
            & $\log(\mathrm{H_2S})$ & -10 -- -1e-6 & [7] \\ \hline
\textbf{Free chemistry}            & $\log(\mathrm{SiO})$ & -10 -- -1e-6 & [8]\\
\textit{high-temperature}           & $\log(\mathrm{TiO})$ & -10 -- -1e-6 & [9]\\
\textit{species}            & $\log(\mathrm{VO})$ & -10 -- -1e-6 & [10]\\
(mass fractions)              & $\log(\mathrm{AlO})$ & -10 -- -1e-6 & [11]\\
            & $\log(\mathrm{SH})$ & -10 -- -1e-6 & [12]\\
\hline
    Temp. Profile & $T$ (K) & 1000 $-$ 3000 \\
\hline
Ref. Radius & $R_\mathrm{ref} ( \mathrm{R_p})$ & 0.8 $-$ 1.2 \\
\hline
Planet gravity  & $\log{g}$  & \multicolumn{2}{l}{$\mathcal{N}(\log{g_\mathrm{p}}, \log{g_\mathrm{p, err}})$} \\

\hline
Clouds & $\log P_\mathrm{cloud} (\mathrm{bar})$ & -8 $-$ 2\\
\hline
Offset &  $\delta_\mathrm{rel} (\mathrm{ppm})$ & -200 $-$ 200 \\ \hline
    \end{tabular}
    
\end{table}

\begin{table}
    \centering
\caption{continued Table\,\ref{tab:ret_priors}, for \nemesispy.}
\label{tab:nemesispy_priors}
    \begin{tabular}{lccc}
         \hline
    \multicolumn{4}{c}{\texttt{NEMESISPY}} \\ \hline
      & \textbf{Parameter}             & \textbf{Prior Range} &  \textbf{Reference}
      \\ \hline

\textbf{Equilibrium}   & $\mathrm{C/O}$  & 0.1 -- 3.9 \\
\textbf{chemistry}  & $\log(\mathrm{M/H})$ & -4 -- 8 \\
\hline
\textbf{Free chemistry}  & $\log(\mathrm{H_2O})$ & -12 -- -0.5 & [1] \\
\textit{base species}            & $\log(\mathrm{CO_2})$ & -12 -- -0.5
 & [2]\\
(VMR)            & $\log(\mathrm{CO})$ & -12 -- -0.5 & [3]\\
& $\log(\mathrm{CH_4})$ & -12 -- -0.5 & [4] \\
            & $\log(\mathrm{NH_3})$ & -12 -- -0.5 & [5]\\
            & $\log(\mathrm{HCN})$ & -12 -- -0.5 & [6] \\
            & $\log(\mathrm{H_2S})$ & -12 -- -0.5 & [7]\\ \hline
\textbf{Free chemistry}            & $\log(\mathrm{SiO})$ & -12 -- -0.5 & [8] \\
\textit{high-temperature}           & $\log(\mathrm{TiO})$ & -12 -- -0.5 & [9]\\
\textit{species}            & $\log(\mathrm{VO})$ & -12 -- -0.5 & [10] \\
(VMR)              & $\log(\mathrm{AlO})$ & -12 -- -0.5 & [11] \\
            & $\log(\mathrm{SH})$ & -12 -- -0.5 & [12]\\
            & $\log(\mathrm{H}^-$) & -15 -- -2 & [13] \\
\hline

\hline

        
6 parameter T-p & $\alpha_1$ & 0.02--2 \\
             & $\alpha_2$ & 0.02--2 \\
             & $\log{P_1}$ & -9 -- 2 \\
             & $\log{P_2}$ & -9 -- 2 \\
             & $\log{P_3}$ & -9 -- 2 \\
             & $T_\mathrm{ref}$ &  1000 -- 3000 \\
\hline

Ref. Radius & $R_\mathrm{ref} ( \mathrm{R_p})$ &  $0.85 - 1.15$ \\
Ref. Pressure  & $P_{\mathrm{ref}}$ (bar) & -9 -- 2  \\

\hline
Clouds  & $\log P_\mathrm{cloud} (\mathrm{bar})$ & -9 -- 2\\
     & $\log{a}$ & -5 -- 5\\
             & $\gamma$ & -10 -- 0 \\
             & $\phi_\mathrm{cloud}$ & 0 -- 1 \\

\hline
Offset &  $\delta_\mathrm{rel}$ (ppm) & $\mathcal{N}(0,500)$ \\ \hline

    \end{tabular}
\end{table}

\section{Retrieval results}
\label{sec:appendix_full_retrievals}
In Table\,\ref{tab:full-retrieval-results}, we show the retrieved parameter from equilibrium and free chemistry retrievals by \pRT and \poseidon using the three reductions and two resolutions of the JWST NIRSpec/G395H retrievals transmission of \myplanet.

\begin{table*}
\def\arraystretch{1.3}
    \caption{Atmospheric retrieval results: the median values with the uncertainties as $16^\mathrm{th}$ and $84^\mathrm{th}$ percentiles. Note that the offset in \poseidon and \pRT is defined in the opposite direction, so their signs are opposite.  }
    \label{tab:full-retrieval-results}
    \begin{tabular}{l c c c c c c c} 
    \hline
     Input spectrum  & $\ln \mathcal{Z}$ & $\log_{10} g$\, (cgs)  &  $T$\,(K) & $R_\mathrm{ref}$\,(R$_\mathrm{Jup}$)  & $\log P_\mathrm{cloud}$\,(bar) & Offset\,(ppm) \\
     \hline
     \eureka, $R=400$   \\
     \pRT: equilibrium chemistry  & $1971.2 \pm 0.1$ & $3.15^{+0.07}_{-0.06}$ & $1620^{+300}_{-320}$ & $1.55\pm 0.02$ & $-4.04^{+1.13}_{-0.83}$ & $-52^{+10}_{-9}$  \\
     \pRT: free chemistry (base) & $1966.3 \pm 0.1$ & $3.11\pm0.08$ & $2040^{+560}_{-610}$ & $1.59 \pm0.01$ & $0.38^{+1.10}_{-1.21}$ & $-66^{+15}_{13}$  \\
     \poseidon: equilibrium chemistry  & $1972.24 \pm 0.11$ & $3.14^{+0.03}_{-0.03}$ & $1380\pm 300$ & $1.54^{+0.01}_{-0.01}$ & $-3.32^{+0.94}_{-1.02}$ & $51^{+10}_{-11}$ \\
     \poseidon: free chemistry (base)  & $1969.14 \pm 0.12$ & $3.13^{+0.03}_{-0.03}$ & $1680^{+630}_{-1060}$ & $1.56^{+0.02}_{-0.01}$ & $0.63^{+0.90}_{-0.94}$ & $67^{+11}_{-13}$\\
    \\ 
    \tiberius, $R=400$ \\
    \pRT: equilibrium chemistry  & $1979.9 \pm 0.1$ & $3.16^{+0.07}_{-0.08}$ & $1350^{+370}_{-270}$ & $1.56^{+0.01}_{-0.02}$ & $-3.44^{+1.15}_{-1.24}$ & $-26\pm 10$  \\
     \pRT: free chemistry (base) & $1976.0 \pm 0.5$ & $3.12^{+0.08}_{-0.07}$ & $1760^{+640}_{-750}$ & $1.57 \pm0.01$ & $0.31^{+1.13}_{-1.31}$ & $-38^{+16}_{-13}$  \\
    \poseidon: equilibrium chemistry  &  $1980.21 \pm 0.11$ & $3.14^{+0.03}_{-0.03}$ & $1220^{+290}_{-180}$ & $1.55^{+0.01}_{-0.02}$ & $-2.89^{+0.67}_{-1.01}$ & $28^{+10}_{-11}$  \\
    \poseidon: free chemistry  (base)  & $1977.52 \pm 0.13$ & $3.12^{+0.03}_{-0.03}$ & $1950^{+510}_{-570}$ & $1.56^{+0.01}_{-0.01}$ & $0.65^{+0.87}_{-0.93}$ & $42^{+12}_{-12}$  \\

    \\
     \exoticjedi, $R=400$\\
    \pRT: equilibrium chemistry  & $1923.8 \pm 0.1$ & $3.17^{+0.07}_{-0.08}$ & $1250^{+270}_{-150}$ & $1.56^{+0.01}_{-0.02}$ & $-2.37^{+2.36}_{-1.24}$ & $53^{+11}_{-10}$  \\
     \pRT: free chemistry (base) & $1918.5 \pm 0.2$ & $3.14\pm 0.08$ & $1310^{+570}_{-430}$ & $1.57 \pm0.01$ & $0.30^{+1.14}_{-1.15}$ & $47^{+19}_{-17}$  \\
    \poseidon: equilibrium chemistry  & $1923.61 \pm 0.12$ & $3.15^{+0.03}_{-0.03}$ & $1130^{+110}_{-70}$ & $1.561^{+0.002}_{-0.005}$ & $-0.60^{+1.80}_{-1.42}$ & $-55^{+11}_{-10}$ \\
    \poseidon: free chemistry (base)  & $1920.83 \pm 0.12$ & $3.14^{+0.03}_{-0.03}$ & $910^{+240}_{-140}$ & $1.570^{+0.004}_{-0.006}$ & $0.41^{+1.04}_{-1.04}$ & $-60^{+14}_{-13}$    \\

     \\
         \eureka, $R=100$ \\
        \pRT: equilibrium chemistry  & $526.4 \pm 0.5$ & $3.14^{+0.06}_{-0.07}$ & $1710^{+280}_{-360}$ & $1.55\pm 0.02$ & $-4.23^{+1.23}_{-0.73}$ & $-54\pm 9$  \\
     \pRT: free chemistry (base) & $520.9 \pm 0.1$ & $3.11\pm0.07$ & $2060^{+560}_{-620}$ & $1.57 \pm0.01$ & $0.41^{+1.04}_{-1.08}$ & $-68^{+14}_{-13}$  \\
     \poseidon: equilibrium chemistry  & $525.33 \pm 0.11$ & $3.13^{+0.03}_{-0.03}$ & $1570^{+380}_{-330}$ & $1.53^{+0.02}_{-0.02}$ & $-3.22^{+0.81}_{-0.96}$ & $54^{+10}_{-9}$ \\
     \poseidon: free chemistry  (base)  & $523.88 \pm 0.12$ & $3.12^{+0.03}_{-0.03}$ & $2120^{+540}_{-650}$ & $1.55^{+0.01}_{-0.01}$ & $0.61^{+0.91}_{-0.94}$ & $69^{+12}_{-13}$  \\
    \\
         \tiberius, $R=100$  \\
         \pRT: equilibrium chemistry  & $531.5 \pm 0.3$ & $3.16\pm 0.06$ & $1560^{310}_{-330}$ & $1.55\pm 0.02$ & $-4.10^{+1.40}_{-0.79}$ & $-28^{+10}_{-9}$  \\
     \pRT: free chemistry (base) & $526.0 \pm 0.1$  & $3.13^{+0.07}_{-0.08}$ & $1870^{+610}_{-770}$ & $1.57 \pm0.01$ & $0.29^{+1.11}_{-1.33}$ & $-37^{+20}_{-14}$  \\
     \poseidon: equilibrium chemistry  & $532.33 \pm 0.11$ & $3.13^{+0.03}_{-0.03}$ & $1430^{+300}_{-270}$ & $1.54^{+0.01}_{-0.02}$ & $-3.31^{+0.89}_{-0.92}$ & $28^{+10}_{-10}$ \\
     \poseidon: free chemistry  (base)  & $529.71\pm0.12$ & $3.13^{+0.03}_{-0.03}$ & $1920^{+550}_{-610}$ & $1.56^{+0.01}_{-0.01}$ & $0.54^{+1.00}_{-1.02}$ & $38^{+13}_{-12}$   \\
    \\
    \exoticjedi, $R=100$ \\
      \pRT: equilibrium chemistry  & $ 501.0 \pm 0.3$ & $3.17\pm 0.07$ & $1420^{+300}_{-270}$ & $1.56\pm 0.02$ & $-3.03^{+1.73}_{-1.16}$ & $44^{+11}_{-10}$  \\
     \pRT: free chemistry (base) & $495.6 \pm 0.1$ & $3.14^{+0.07}_{-0.08}$ & $1390^{+740}_{-510}$ & $1.57 \pm0.01$ & $0.34^{+1.10}_{-1.23}$ & $39^{+20}_{-17}$  \\
     \poseidon: equilibrium chemistry  & $501.70 \pm 0.11$ & $3.14^{+0.03}_{-0.03}$ & $1210^{+290}_{-120}$ & $1.56^{+0.01}_{-0.02}$ & $-1.97^{+2.30}_{-0.96}$ & $-44^{+11}_{-11}$ \\
     \poseidon: free chemistry (base)  & $500.12 \pm 0.12$ & $3.13^{+0.03}_{-0.03}$ & $1320^{+640}_{-490}$ & $1.56^{+0.01}_{-0.01}$ & $0.49^{+1.01}_{-1.06}$ & $-43^{+15}_{-17}$ \\
     \hline
      
    \end{tabular}

\end{table*}

\section{Posterior distributions} 
\label{sec:appendix_posterior_plots}
\begin{figure*}
    \centering
    \includegraphics[width=\linewidth]{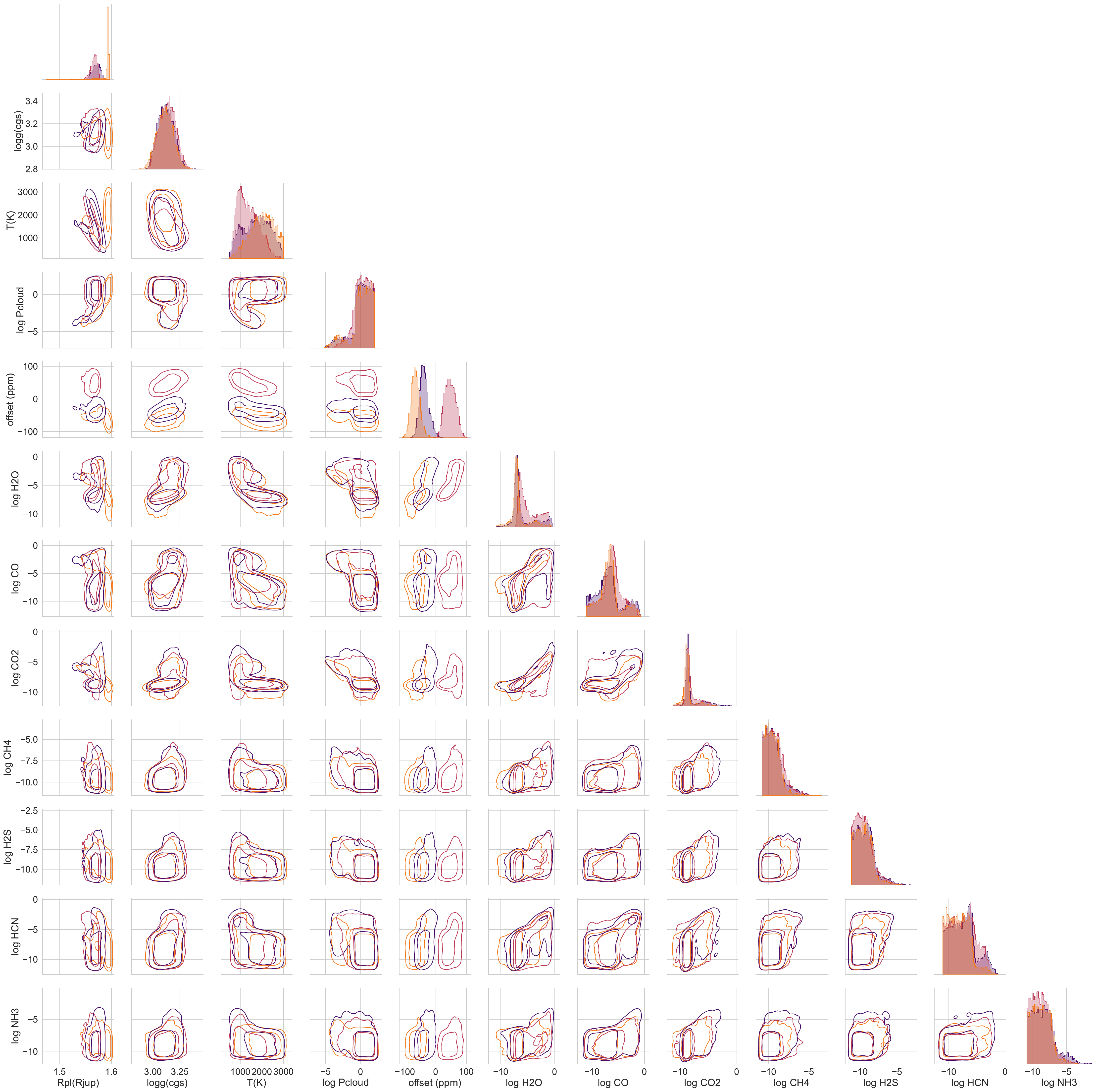}
    \caption{Retrieved posterior distributions using \pRT with the free chemistry setup for the three reductions at R=400.  }
    \label{fig:corner-free-chem-r400-nirspec-only}
\end{figure*}

\begin{figure*}
    \centering
    \includegraphics[width=\linewidth]{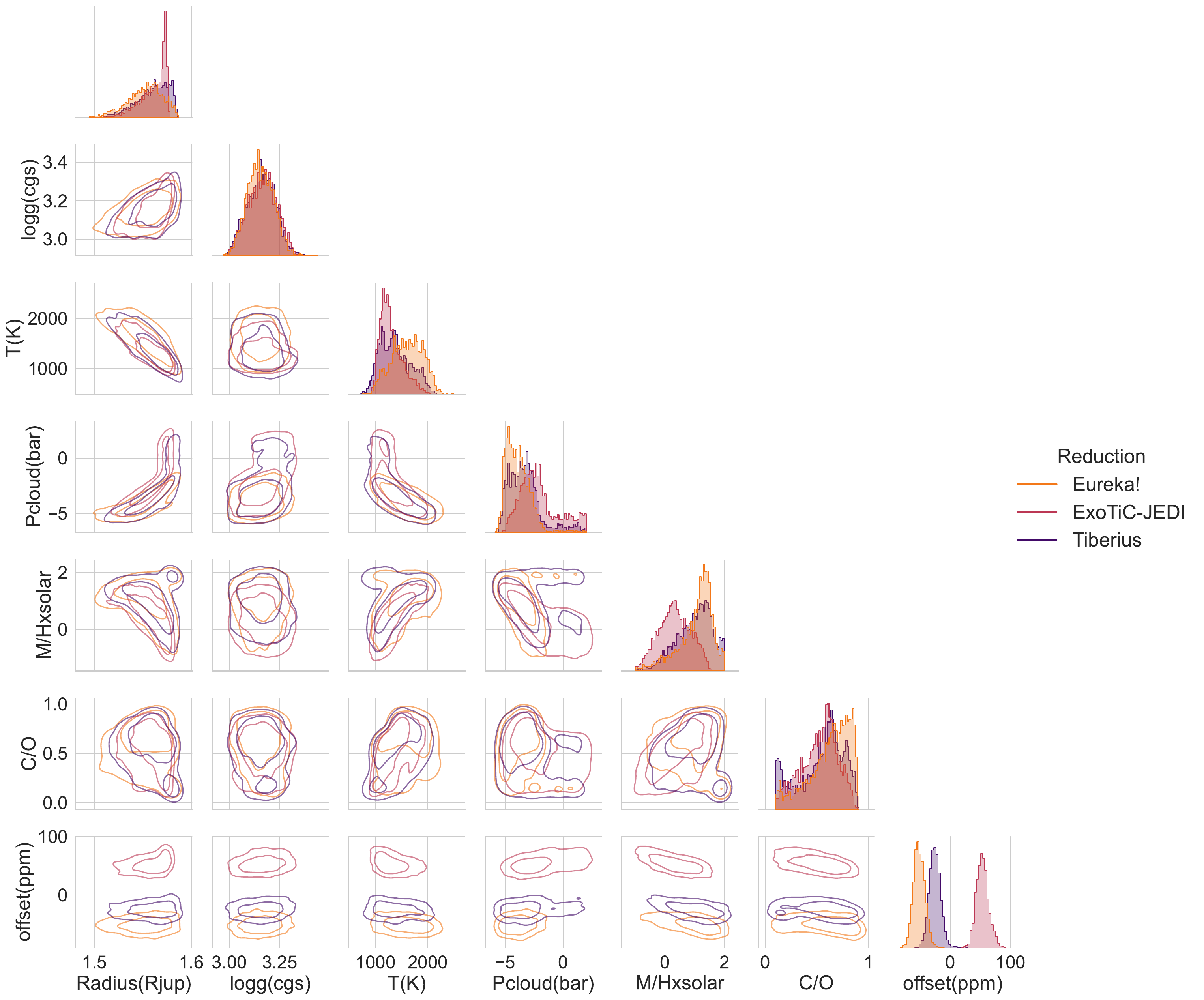}
    \caption{Retrieved posterior distributions using \pRT with the equilibrium chemistry setup for the three reductions at R=400.  }
    \label{fig:corner-eq-chem-r400-nirspec-only}
\end{figure*}

Here we show additional posterior distributions of our retrievals using the three reductions at R=400. In Fig.\,\ref{fig:corner-eq-chem-r400-nirspec-only} we display the equilibrium chemistry corner plot while Fig.\,\ref{fig:corner-free-chem-r400-nirspec-only} shows the free chemistry setup, both using \pRT.

\begin{figure*}
    \centering
    \includegraphics[width=\linewidth]{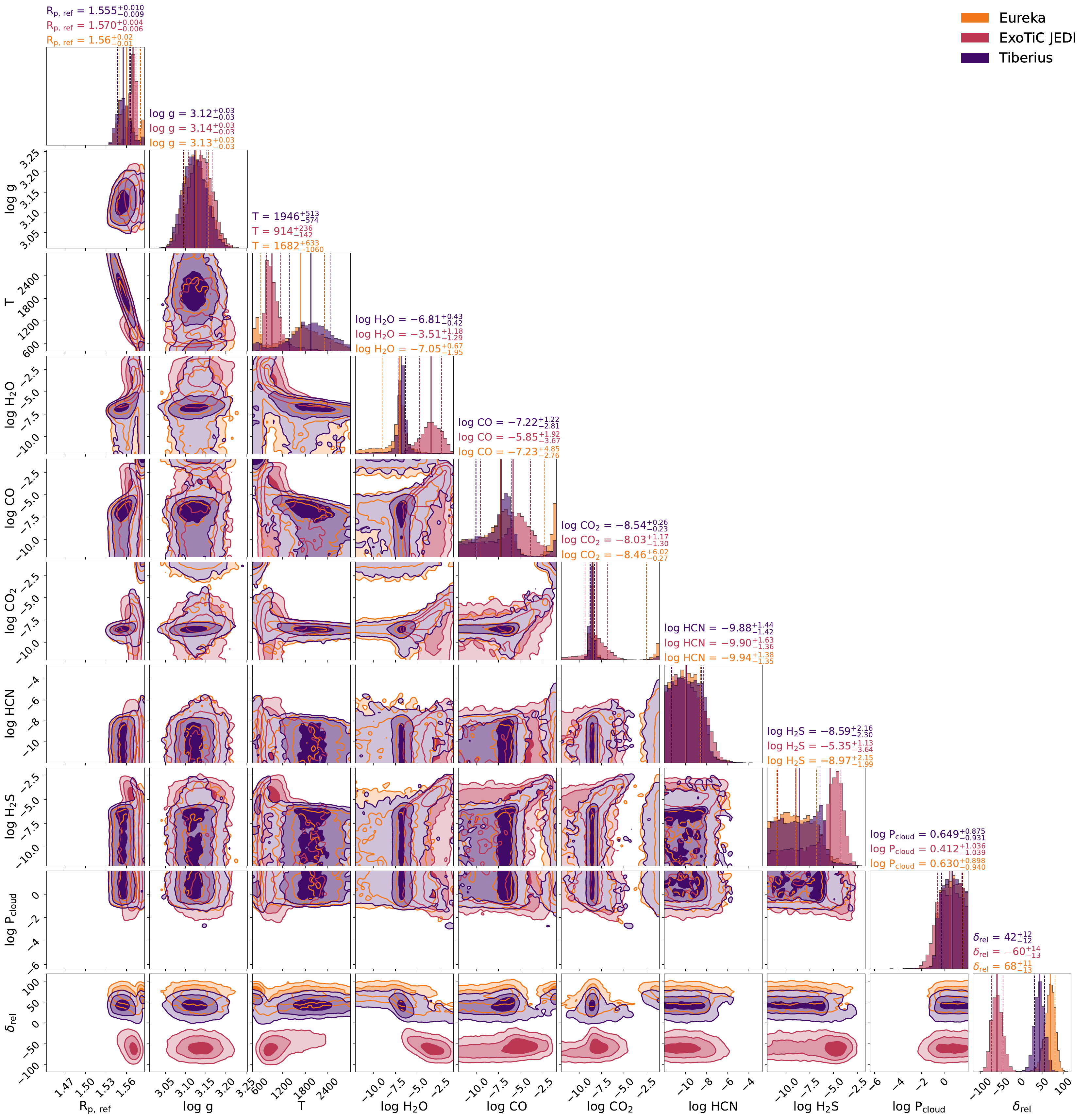}
    \caption{Retrieved posterior distributions using \poseidon with the free chemistry setup for the three reductions at R=400.  }
    \label{fig:corner-free-chem-r400-nirspec-only-pos}
\end{figure*}

\begin{figure*}
    \centering
    \includegraphics[width=\linewidth]{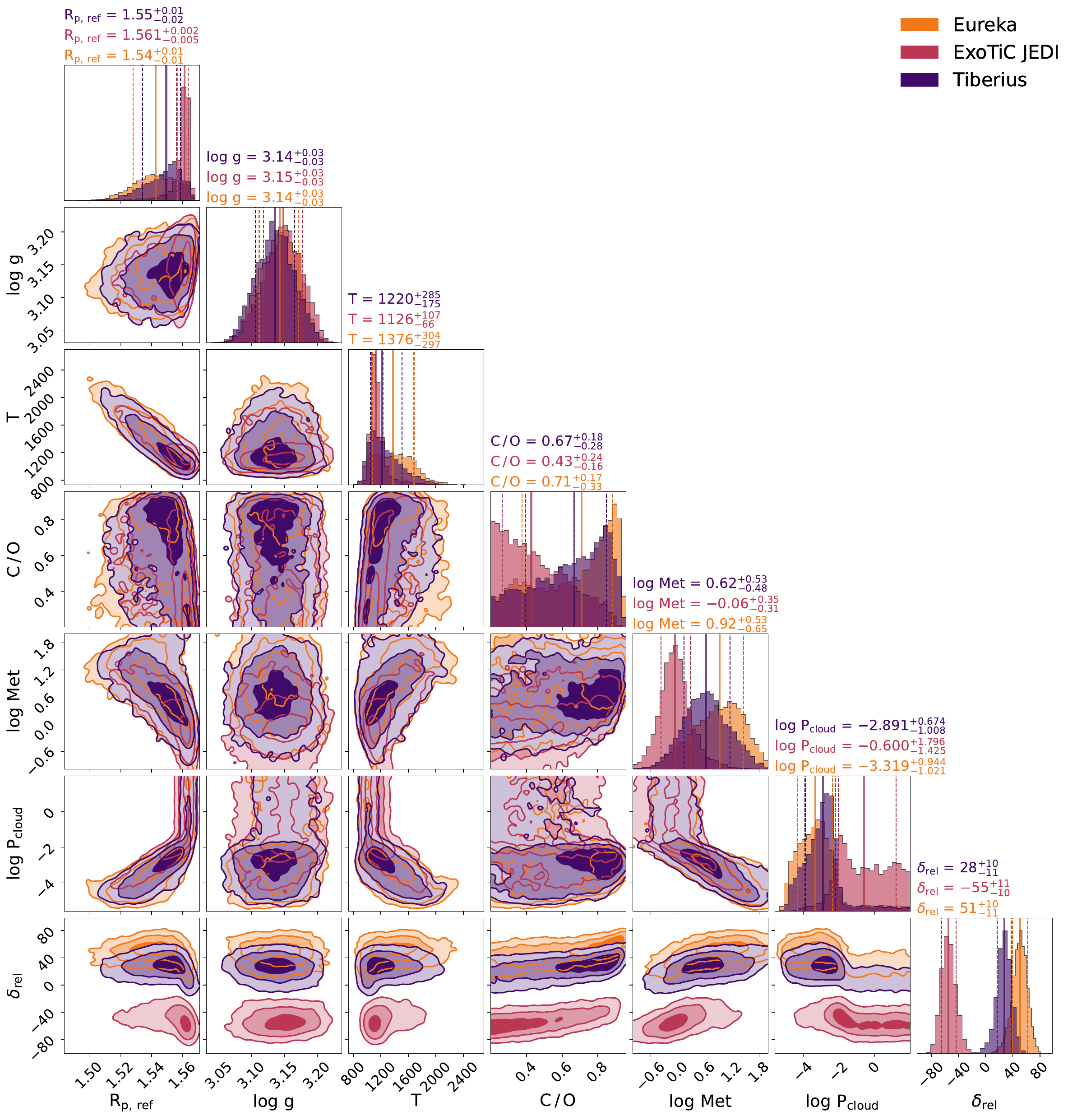}
    \caption{Retrieved posterior distributions using \poseidon with the equilibrium chemistry setup for the three reductions at R=400.  }
    \label{fig:corner-eq-chem-r400-nirspec-only-pos}
\end{figure*}

\section{Combined JWST + HST equilibrium chemistry retrieval}
\begin{figure*}
    \centering
    \includegraphics[width=\textwidth]{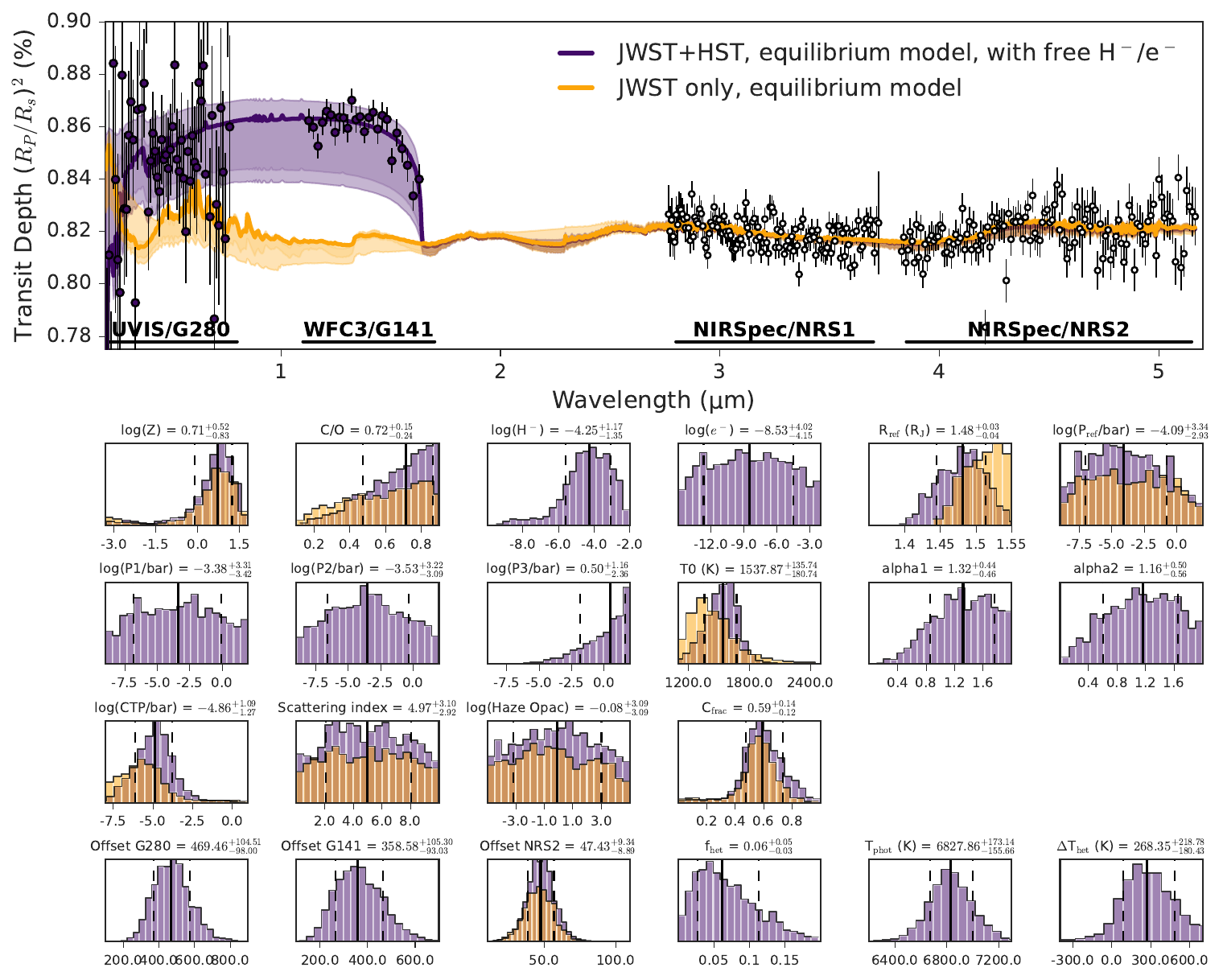}
    \caption{Top: NEMESISPY equilibrium chemistry retrievals comparison between using JWST data in combination with previously published HST data (purple: including \ce{H-} and \ce{e-}) and JWST data only (orange). The colour of the data corresponds to the offset applied from each model. Whilst the retrieval including the HST data was conducted with the G280 dataset held fixed, here we shift the plot such that the offsets are shown relative to NRS1 to facilitate comparison with figure~\ref{fig:appendix_g395h_hst}. \newline Bottom: The posterior plots for the metallicity and C:O ratio, H$^-$ and e$^-$, R$_{\mathrm{ref}}$ and P$_{\mathrm{ref}}$; temperature (P1,P2,P2,T0,alpha1,alpha2); cloud properties (cloud top pressure, haze scattering index, haze opacity and cloud fraction); offsets; and stellar parameters (heterogeneity fraction, photospheric temperature, heterogeneity temperature relative to photosphere).  The colours correspond to the individual models, demonstrating that the inclusion of HST data does not provide further constraints for the cloud deck or metallicity. The discrepancy in the retrieved radius is due to the two retrievals referencing the instrument offsets to different baseline instruments (G280 for the full retrieval and NRS1 for the JWST-only retrieval). The offsets for both retrievals are presented relative to NRS1 to aid comparison.}.
    \label{fig:appendix_g395h_hst_nemesispy}
\end{figure*}

Here we show the \nemesispy equilibrium chemistry retrieval with free \ce{H-} and \ce{e-} abundance results when combining our JWST NIRSpec/G395H with previously published HST transmission spectra of \myplanet, Fig.\,\ref{fig:appendix_g395h_hst_nemesispy}.

\bsp	
\label{lastpage}
\end{document}